\documentclass[prd,superscriptaddress,floatfix,notitlepage,nofootinbib,reprint,letterpaper]{revtex4-1} 
\pdfoutput=1
\usepackage{graphicx}
\usepackage{hyperref}
\usepackage{color}
\usepackage{amsmath,amssymb}
\usepackage{epsfig}
\usepackage{bm}
\usepackage{natbib,ifthen}
\usepackage{wasysym}
\usepackage{mathrsfs}

\newcommand{\comment}[1]{}

\newcommand{\beq}{\begin{equation}}
\newcommand{\eeq}{\end{equation}}
\newcommand{\bea}{\begin{eqnarray}}
\newcommand{\eea}{\end{eqnarray}}
\newcommand{\bsp}{\begin{split}}
\newcommand{\esp}{\end{split}}

\newcommand{\hMpc}{\ h^{-1}\text{Mpc}}

\renewcommand{\vec}[1]{\bm{#1}}

\newcommand{\vx}{\vec x}
\newcommand{\vv}{\vec v}
\newcommand{\vs}{\vec s}
\newcommand{\vk} {{\boldsymbol k}}
\newcommand{\vp} {{\boldsymbol p}}
\newcommand{\vq} {{\boldsymbol q}}
\newcommand{\vps}{\vec \psi}
\newcommand{\vc}{\vec \chi}

\newcommand{\vchi}{\vec \chi}
\newcommand{\vnabla}{{\boldsymbol \nabla}}

\newcommand{\ihMpc}{\; h\text{Mpc}^{-1}}

\newcommand{\la}{\left\langle}
\newcommand{\ra}{\right\rangle}

\definecolor{darkgreen}{RGB}{0,120,0}

\newcommand{\eq}[1]{(\ref{eq:#1})} 
\newcommand{\eqq}[1]{Eq.~(\ref{eq:#1})} 
\newcommand{\fig}[1]{Fig.~\ref{fig:#1}} 
\newcommand{\se}[1]{Section~\ref{se:#1}} 
\newcommand{\app}[1]{Appendix~\ref{app:#1}}

\newcommand{\mnras}{Mon.\ Not.\ R.\ Astron.\ Soc.}

\newcommand{\jcap}{JCAP}
\newcommand{\apjl}{APJL~}
\newcommand{\physrep}{Physics Reports}
\newcommand{\aap}{Astronomy \& Astrophysics}

\begin{document}
\title{Iterative initial condition reconstruction}

\author{Marcel Schmittfull}
\affiliation{Institute for Advanced Study, Einstein Drive, Princeton, NJ 08540, USA}
\author{Tobias Baldauf}
\affiliation{Centre for Theoretical Cosmology, DAMTP, University of Cambridge, CB3 0WA, UK}
\author{Matias Zaldarriaga}
\affiliation{Institute for Advanced Study, Einstein Drive, Princeton, NJ 08540, USA}

\date{\today}

\begin{abstract}
Motivated by recent developments in perturbative calculations of the nonlinear evolution of large-scale structure, we present an iterative algorithm to reconstruct the initial conditions in a given volume starting from the dark matter distribution in real space. 
In our algorithm, objects are first moved back iteratively along estimated potential gradients, with a progressively reduced smoothing scale, until a nearly uniform catalog is obtained.
The linear initial density is then estimated as the divergence of the cumulative displacement, with an optional second-order correction.
This algorithm should undo nonlinear effects up to one-loop order, including the higher-order infrared resummation piece.
We test the method using dark matter simulations in real space.
At redshift $z=0$, we find that after eight iterations the reconstructed density is more than $95\%$ correlated with the initial density at $k\le 0.35\; h\mathrm{Mpc}^{-1}$.
The reconstruction also reduces the power in the difference between reconstructed and initial fields by more than 2 orders of magnitude at $k\le 0.2\; h\mathrm{Mpc}^{-1}$, and it extends the range of scales where the full broadband shape of the power spectrum matches linear theory by a factor of 2-3.
As a specific application, we consider measurements of the baryonic acoustic oscillation (BAO) scale that can be improved by reducing the degradation effects of large-scale flows.  
In our idealized dark matter simulations, the method improves the BAO signal-to-noise ratio by a factor of 2.7 at $z=0$ and by a factor of 2.5 at $z=0.6$, improving standard BAO reconstruction by $70\%$ at $z=0$ and $30\%$ at $z=0.6$, and matching the optimal BAO signal and signal-to-noise ratio of the linear density in the same volume.
For BAO, the iterative nature of the reconstruction is the most important aspect. 
\end{abstract}

\maketitle

\section{Introduction}

Studies of large scale structure (LSS) and the cosmic microwave background (CMB) have played a very important role in establishing the standard model of cosmology. Successful CMB experiments over the past few decades have managed to extract almost all of the available information in the primary anisotropies, and thus the importance of LSS studies is expected to increase in the future. Because of the sheer number of independent modes that can be measured in our observable Universe, in principle three-dimensional maps of the late Universe contain a vast amount of statistical information about the initial conditions for structure formation as well as cosmological parameters. This is especially true if one can extract information from small spatial scales. 

The fluctuations we observe on small scales in the late Universe have been heavily processed by their dynamical evolution progressively scrambling the cosmological information to the extent that sufficiently small scales are usually excluded from cosmological analysis. The problem is made even worse by the fact that many of the remaining open questions in cosmology such as measuring the masses of neutrinos or constraining the Gaussianity of the initial seeds require measuring very small effects. 

Thus there is great motivation for and a long history of trying to undo the cosmological evolution in a process usually called reconstruction. The hope is to unscramble the cosmological information that might still be present in higher order moments of the data and recover the initial conditions directly which can then be described using a power spectrum and perhaps a few low-order statistics. These techniques have been used most successfully in the context of baryonic acoustic oscillation (BAO) measurements. The BAO feature in the correlation function acts like a standard ruler whose physical size is calibrated by CMB measurements, so that measuring its angular extent on the sky as a function of redshift measures the expansion history of the Universe.
However, the BAO feature is degraded (i.e.~it gets broader) as a result of nonlinear evolution. The process of reconstruction sharpens back the peak, increasing the signal-to-noise ratio.

The operational procedure of reconstruction methods is usually motivated by analytical models that connect the linear initial density with the observed nonlinear density in the perturbative regime. In the past few years a lot of work has been put into trying to find improved models; for example a recent discussion of Lagrangian-space models can be found in  \cite{Matsubara:2007wj,Matsubara0807,2014PhRvD..90d3537M,2014JCAP...05..022P,Carlson2013,zvonimir1410,2015PhRvD..91l3516S,Vlah:2015sea,Vlah:2016,2016JCAP...01..043M,Baldauf:2015tla,Baldauf:2015zga}. Here we are motivated by developments connected with the effective field theory (EFT) approach to large-scale structure \cite{2012JCAP...07..051B,2012JHEP...09..082C,2014PhRvD..90b3518C,SenatoreZaldarriaga1404} and  ask whether these theoretical modeling advances can be used to develop an improved reconstruction method. This is the goal of our paper, where we will focus on a Lagrangian EFT model developed in \cite{Baldauf:2015zga}.

Related to the motivation for standard reconstruction \cite{EisensteinRec}, the key motivation for our method is that the dominant nonlinear terms that reduce the correlation of the nonlinear density with initial conditions and degrade the BAO scale are nonlinear shift terms (rather than intrinsic nonlinearities like nonlinear growth or tidal terms).  
These shift terms can be reduced if we have a good estimate of the nonlinear displacement field from initial Lagrangian to final Eulerian coordinates. 
Motivated by \cite{Baldauf:2015zga}, we estimate the nonlinear displacement field by iteratively moving particles back along estimated Zeldovich displacements.
The resulting cumulative displacement field significantly improves the single-step Zeldovich displacement used in standard reconstruction.
This is the most important aspect of our method.
We refine the method further by adding second order corrections when estimating the linear density from the nonlinear displacement field, but this is less important for most applications.
For simplicity we will only consider the idealized toy model of dark matter in real space in this paper.
In future work we plan to extend the method to halos in redshift space, accounting also for realistic shot noise and survey selection functions.

The subject of reconstruction was pioneered in Refs.~\cite{1989ApJ...344L..53P,1990ApJ...362....1P}, where the initial positions of eight Local Group galaxies were estimated by minimizing the action of trial orbits assuming negligible initial peculiar velocities,
and in Ref.~\cite{1989ApJ...336L...5B}, where the assumption of irrotational Lagrangian velocities and the Zeldovich approximation \cite{Zeldovich1970} were used to estimate the linear velocity potential and linear mass density from observed radial peculiar velocities.
In a different approach called Gaussianization, the nonlinear density was transformed in a rank-preserving way such that its one-point probability distribution function becomes Gaussian \cite{1992MNRAS.254..315W}.
In a method more similar to most current methods, including ours, the Zeldovich approximation was used to describe the time evolution of the velocity potential that can be integrated back in time without generating vorticity \cite{1992ApJ...391..443N}.
Mass conservation can be imposed using the continuity equation \cite{1993ApJ...405..449G}, and an iterative higher-order scheme related to this was proposed in Ref.~\cite{1999MNRAS.308..763M}.
In the path interchange Zeldovich approximation (PIZA) method \cite{1997MNRAS.285..793C} the least action principle was applied assuming trajectories are straight lines, which can be combined with the Gaussianization method \cite{1999ApJ...515..471N}, or generalized  beyond straight trajectories \cite{2000MNRAS.313..587N,2002MNRAS.335...53B}.  
More recently, this was generalized to more realistic dynamics, finding that different methods perform similarly for velocity reconstruction if applied to realistic mock catalogs \cite{KeselmanNusser1609}.
In another generalization known as Monge-Ampere-Kantorovich reconstruction, an optimal mass assignment problem is solved \cite{2002Natur.417..260F,2003MNRAS.346..501B,2006MNRAS.365..939M,SirkoThesis,2008PhyD..237.2145M}.

A related method now referred to as standard reconstruction \cite{EisensteinRec} first computes the Zeldovich displacement from the filtered nonlinear density (corresponding to the first step of the iterative procedure of our method), and then moves the clustered and a random catalog by that displacement to estimate the linear density from the density difference of the two displaced catalogs (this procedure removes the nonlinear shift terms that otherwise degrade the BAO scale, also see \cite{Padmanabhan0812,TassevRec}).
That method was successfully applied to real galaxy survey data from SDSS BOSS \cite{Padmanabhan2012BAORec,AndersonBAODR9_1203,1409.3242,Anderson1312.4877,Florian1506.03900,BOSSFinalAlam} and WiggleZ \cite{2014MNRAS.441.3524K}, improving BAO measurements typically by a factor of $\sim 2$.
Similar improvements are expected for future galaxy redshift surveys including, for example, DESI \cite{DESIwhitepaper}, Euclid \cite{EuclidWhitePaper}, and LSST \cite{LSSTSciBook0912}.

A conceptually different approach to reconstruction is to sample the initial condition density, propagate it forward using an $N$-body or 2LPT code, compare against observations, resample modes that did not fare well, etc, which becomes computationally feasible using Hamiltonian sampling techniques \cite{JascheWandelt1203,2013MNRAS.429L..84K}.
Another method to reconstruct the initial density from the observed density using the Hamiltonian MCMC technique  was presented in \cite{2013ApJ...772...63W,2009MNRAS.394..398W}.
While these methods fix the initial linear power spectrum and cosmology to constrain the initial modes, a converse approach \cite{SeljakEtAlRec} is to marginalize over the modes and constrain the initial power spectrum and cosmology using optimization techniques.

\begin{table*}[thbp]
\centering
\renewcommand{\arraystretch}{1.4}
\begin{tabular}{@{}llp{6.5cm}lp{6cm}@{}}
\toprule
Reconstruction algorithm &\phantom{}&  Stage I: Estimate displacement &\phantom{}& Stage II: Estimate linear density \\
\colrule
Standard reconstruction \cite{EisensteinRec} && Zeldovich $\hat\vchi_\mathrm{ZA}=\frac{i\vk}{k^2}W\delta_\mathrm{NL}$ && $\hat\delta_0=\delta_d[\hat\vchi_\mathrm{ZA}]-\delta_s[\hat\vchi_\mathrm{ZA}]$ \\
Iterative standard rec \cite{SeoBAOShift} && Iterative Zeldovich with fixed $W$ && $\hat\delta_0=\delta_d[\hat\vchi_\mathrm{ZA}]-\delta_s[\hat\vchi_\mathrm{ZA}]$ \\
Improved standard rec \cite{TassevRec} && Zeldovich and iterative Newton-Raphson && $\hat\delta_0=\hat\delta_\chi=\vnabla\cdot\hat\vchi$ \\
Standard rec with pixels \cite{SeoHirataPixelRec1508,ObuljenPixelRec1610} && Zeldovich $\hat\vchi_\mathrm{ZA}=\frac{i\vk}{k^2}W\delta_\mathrm{NL}$ && Move pixels instead of galaxies\\
Eulerian growth-shift rec \cite{Marcel1508} && Zeldovich $\hat\vchi_\mathrm{ZA}=\frac{i\vk}{k^2}W\delta_\mathrm{NL}$ && $\hat\delta_0=\delta_\mathrm{NL}-\hat\vchi_\mathrm{ZA}\cdot\vnabla\delta-\delta^2$ \\
Nonlinear isobaric rec \cite{IsoRecZhu1609OneD,IsoRecZhu1611ThreeD} && Solve diff.~eqn.~with multigrid algorithm && $\hat\delta_0=\hat\delta_\chi=\vnabla\cdot\hat\vchi$ \\
New $\mathcal{O}(1)$ rec && Iteratively solve $T(k)F_Z[\vnabla\!\cdot\!\vchi]=W\delta_\mathrm{NL}$ for $\vchi$ && $\hat\delta_0=\hat\delta_\chi=\vnabla\cdot\hat\vchi$ \\
New $\mathcal{O}(2)$ rec && Iteratively solve $T(k)F_Z[\vnabla\!\cdot\!\vchi]=W\delta_\mathrm{NL}$ for $\vchi$ && $\hat\delta_0=t_1(k)\hat\delta_\chi+t_2(k)\int_\vp\kappa_2\hat\delta_\chi(\vp)\hat\delta_\chi(\vk-\vp)$ \\
Extended std rec [appdx.~\ref{app:ExtendedStdRec}] && Iteratively solve $T(k)F_Z[\vnabla\!\cdot\!\vchi]=W\delta_\mathrm{NL}$ for $\vchi$ && $\hat\delta_0=\delta_d[\hat\vchi]-\delta_s[\hat\vchi]$ \\
\botrule
\end{tabular}
\caption{Overview of some recently proposed BAO reconstruction techniques that estimate first a displacement field from initial to final conditions, and then use that to infer the linear density.
The methods in the bottom three rows are new to this paper, and a crucial ingredient separating them from other methods is that the smoothing scale is progressively reduced with the number of iteration steps to access smaller scales.
The list is by no means complete; see main text for more methods.
}
\label{tab:CompareAlgorithms}
\end{table*}

Following the success of the standard reconstruction method \cite{EisensteinRec} for improving measurements of the acoustic peak, a series of related reconstruction procedures were proposed recently to achieve further improvements or enable broader applications beyond BAO measurements \cite{SeoBAOShift,TassevRec,SeoHirataPixelRec1508,ObuljenPixelRec1610,Marcel1508,IsoRecZhu1609OneD,IsoRecZhu1611ThreeD}.
These methods have a common structure: First, they estimate the displacement field from initial to final positions, and second, they estimate the linear density from that displacement.
However, the concrete procedure for these two stages differs for the various proposed methods as summarized in Table~\ref{tab:CompareAlgorithms}.
The practical application of standard reconstruction to real galaxy survey data and improved implementations are discussed, for example, in \cite{Padmanabhan2012BAORec,Burden1408,Achitouv1507,Burden1508}.
In our method described below, we will modify both the first and second stage based on a concrete analytical model for structure formation.

An earlier similar attempt to improve the displacement field using iterative Zeldovich displacements and higher-order corrections \cite{SeoBAOShift} found only negligible improvements over standard reconstruction.
We attribute the main reason for this to the fact that the filtering smoothing scale was kept fixed in the iterative procedure, so that only shifts on very large scales were mitigated.  Among other changes and differences in theoretical motivations, we improve that method by progressively reducing the smoothing scale with each iteration step (also see \app{ExtendedStdRec}).

Recently, a related reconstruction method was proposed in a series of papers by Zhu \emph{et al.} \cite{IsoRecZhu1609OneD,IsoRecZhu1611ThreeD}. 
This method is also based on improved estimates of the nonlinear displacement between initial Lagrangian and final Eulerian coordinates, finding substantial improvements over standard reconstruction in terms of the cross-correlation with the linear density \cite{IsoRecZhu1609OneD,IsoRecZhu1611ThreeD}, Fisher information \cite{IsoRecPan1611Fisher}, BAO \cite{IsoRecWang1703BAO}, as well as when applying it to halos \cite{IsoRecYu1703Halos}. 
In that approach, the displacement is estimated by writing the deformation of the observed density grid to an initially uniform grid under the continuity equation as a differential equation that is then solved numerically using a multigrid moving mesh algorithm. 
Where comparisons are possible, this method seems to perform very similarly to ours.
Despite differences in the theoretical motivation and operational procedure for constructing the displacement field and for inferring the linear density from that displacement, the similarity in final performance is likely related to the fact that both methods rely on significantly improved displacement fields compared to standard reconstruction.
We believe both approaches shall be useful in the future to improve BAO measurements and realize other applications of reconstruction.

Our paper is structured as follows. We start in \se{TheoryMotivation} with the theoretical motivation for the particular reconstruction algorithm that we propose. \se{NumericalSetup} describes the implementation of the algorithm and the numerical setup and simulations.
In \se{Results} we assess the performance of the reconstruction based on the one-point probability distribution function (pdf) of the reconstructed density, its cross-correlation with the linear initial conditions, the BAO signature in the power spectrum, and the full shape of the power spectrum.
We conclude in \se{Conclusions}. 
\app{TransferFcns} discusses transfer functions that are needed when including second-order corrections to the reconstruction.
In \app{ExtendedStdRec} we describe a simple extension of standard reconstruction that performs worse than the method described in the main text, but still much better than standard reconstruction.  
\app{BAOFitting} describes our procedure to fit the BAO signature in the power spectrum.
\app{HighZ} and \app{TechnicalDetails} discuss results at higher redshift, the choice of reconstruction parameters, and simulation convergence tests.
\app{RecModel} provides a perturbative analysis motivating the reconstruction algorithm further.

\section{Theoretical motivation}
\label{se:TheoryMotivation}

Perturbation theory has been successful in modeling the large-scale statistical properties of the evolved LSS given the initial conditions. In the past few years the application of effective field theory (EFT) ideas has allowed additional improvements on scales larger than the nonlinear scale. Regardless of these improvements, perturbation theory breaks down on small scales once shell crossing has occurred. In addition,  we also expect reconstruction algorithms to break in the same regime as it is not possible to uniquely infer the initial density field from the final density once  shells have crossed. One could think of a shell of matter of some radius; given just the density one cannot know if the shell is collapsing for the first time or moving back out. Equivalently one could imagine finding a solution for the displacement of a uniform set of particles that after being displaced lead to an observed density field. Given any such solution, one can easily find another one by swapping particles  around.  Thus if perturbation theory and any reconstruction algorithm are expected to fail on the same scales, it seems relevant to try to use our better understanding of perturbation theory to improve reconstruction, as an algorithm based on perturbation theory might do as well as one possibly can do. We will take the first steps along this direction in this paper.
 
In particular, although BAO reconstruction has been very successful, we know it does not recover all the information that is there in the linear field. Thus, if we manage to improve the reconstruction algorithm, we might expect (or hope)  that for sufficiently good data there might be some room for improvement in BAO related measurements.  There is also the hope that reconstruction might be useful for studying other questions.

Let us start by being more precise about our theoretical expectations:
\begin{itemize}
\item Perturbation theory only works on large scales, so we should only try to use the large scale nonlinear density as an input to the reconstruction algorithm. 
\item In perturbation theory there are two sources of nonlinearities: shift terms and true changes to the small scale dynamics, for example corrections to the growth rates. These terms are of different sizes. We see this directly when we compare one-loop corrections to the displacement and density fields (the displacement is a Lagrangian quantity that is not corrected by the shift terms).  The scale at which the one-loop correction to the displacement power is equal to linear displacement power is much smaller (at higher $k$) than that for which the one-loop corrections to the density power are equal to the linear density power (e.g.~\cite{Baldauf:2015tla,Baldauf:2015zga}). 
This shows that the biggest effect in the broadband error of linear theory is from shift terms, which affect the density but not the displacement. We discuss this more quantitatively in \app{RecModel}.
\item If we are interested in the BAO feature, we also know that it is the shift terms that damp the oscillations (broaden the peaks) \cite{EisensteinSeoWhite0604,EisensteinRec}.  
\item We also know that the correlation function around the BAO scale computed in the Zeldovich approximation is almost the same as the fully nonlinear one. It is very difficult to spot the additional corrections from the other one-loop terms (i.e.~the one-loop IR-resummed correlation function is difficult to tell apart from Zeldovich). Thus we expect that undoing Zeldovich is almost all that is necessary.  
\item If one assumes just one stream, there is a unique way to invert Zeldovich.  But once shell crossing has occurred, no unique solution exists, and the problem is no longer invertible. So on small scales the displacement inferred by making the density uniform (inverting Zeldovich) cannot be related to the initial conditions in a simple form.  Finding the correct displacement might not even be possible with simulations rather than perturbation theory.  At best one can rely on a prior to rank the probability of different solutions.   
 \end{itemize}

These considerations suggest estimating the perturbative part of the density field by filtering the nonlinear density field, then inverting the relation between displacement and density to obtain the displacement that would transform a uniform density to the filtered nonlinear density, and then estimate the linear density from that displacement.

To see this more quantitatively, we can start from the Lagrangian model built in \cite{Baldauf:2015zga}
\begin{align}
\label{eq:deltaNLmodel}
\delta_\mathrm{NL} = T(k) \delta_Z[\delta_\chi] + \delta_{error} \equiv \delta_{PT} + \delta_{error},
\end{align}
where
\begin{align}
\label{eq:deltaChiAsFcnOfDelta0}
\delta_\chi (\vk)&= a_1(k) \delta_0(\vk) + a_2(k) \int {d^3\vp_1 \over (2\pi)^3}\  {3\over 14}\nonumber\\
&\quad\;\times \left(1-{(\vp_1\cdot\vp_2)^2 \over \vp_1\cdot\vp_1 \vp_2\cdot\vp_2 }\right)  \delta_0(\vp_1) \delta_0(\vp_2) + \cdots.
\end{align}
Here $\delta_\mathrm{NL}$ is the nonlinear final density and $\delta_0$ is the linear initial density.
$a_1(k)$, $a_2(k)$ and $T(k)$ are transfer functions that only depend on the modulus of the Fourier wave vector.
$\delta_Z[\delta_\chi]$ is the density obtained by moving a uniform set of particles by a displacement $\vchi$ whose divergence is $\delta_\chi\equiv\vnabla\cdot\vchi$. This model contains all the terms relevant for a one-loop calculation. In principle one should add a cubic term but up to one loop only the part that correlates with the linear density enters. Thus the cubic term is included in the $a_1$ transfer function. In the EFT approach one has to include counterterms, which at this order are described by the so-called speed of sound and related terms. These contributions are absorbed in $a_1$ and $T$. Finally, since we are working in Lagrangian space, all shift terms are included to all orders. Thus this scheme already does the so-called infrared resummation. 

One should think of $\vchi$ as the nonlinear displacement from initial Lagrangian coordinates to late-time Eulerian coordinates, which can be modeled perturbatively in the initial (linear) density field $\delta_0$ as in \eqq{deltaChiAsFcnOfDelta0}, with $\vk=\vp_1+\vp_2$.
The error density $\delta_\mathrm{error}$ accounts for the difference between the perturbatively based prediction and the simulation results.  Its power spectrum was presented in \cite{Baldauf:2015zga}. That paper showed that not much is gained by going to higher order in perturbation theory for $\delta_\chi$. Even with the quadratic based prediction for $\delta_\chi$, the cross-correlation coefficient between the perturbative and nonlinear fields drops to 0.5 only at $k\approx 0.5\ihMpc$ at redshift $z=0$.

Assuming the model of Eqs.~\eq{deltaNLmodel} and \eq{deltaChiAsFcnOfDelta0} is the correct connection between linear density $\delta_0$ and nonlinear density $\delta_\mathrm{NL}$, how can we estimate $\delta_0$ from the observed $\delta_\mathrm{NL}$?
\eqq{deltaNLmodel} suggests that we first want to estimate the perturbative part of the density, $\delta_{PT}$, from the nonlinear density field. If all fields were Gaussian, that would involve a simple Wiener filter. We will follow the same procedure here and estimate $\delta_{PT}$ by filtering the nonlinear density field,
\beq
\hat \delta_{PT}(\vk)= W(k) \delta_\mathrm{NL}(\vk), 
\eeq
where $W$ is a filter which we choose to be a Gaussian for simplicity. 
Then we estimate the divergence $\delta_\chi$ of the Lagrangian-to-Eulerian displacement by demanding
\beq
\label{eq:DisplIsFilteredNonl}
T(k) \delta_Z[\hat \delta_\chi] =  \hat \delta_{PT} = W \delta_\mathrm{NL}.
\eeq
This is a nonlinear equation for $\hat \delta_\chi$, which we will solve iteratively.
The solution is an estimate $\hat\vchi$ of the Lagrangian-to-Eulerian displacement, and represents the first stage of the reconstruction.
In a second stage, we estimate the linear density $\hat\delta_0$ corresponding to that displacement $\hat\vchi$.
If $\hat\vchi$ contained only linear displacements, the linear density would simply be given by its divergence.
If $\hat\vchi$ also contains nonlinear displacements, we can model the relation between nonlinear displacement and linear density using \eqq{deltaChiAsFcnOfDelta0},  and invert that to estimate the linear density from the nonlinear displacement using
\begin{align}
\label{eq:hat_delta0_from_2ndorder}
\hat \delta_0 (\vk) & = t_1(k) \hat\delta_\chi(\vk) 
+ t_2(k) \int {d^3\vp_1 \over (2\pi)^3}\  
 \nonumber\\
&\quad\;\times  \left(1-{(\vp_1\cdot\vp_2)^2 \over \vp_1\cdot\vp_1 \vp_2\cdot\vp_2 }\right) \hat \delta_\chi(\vp_1) \hat \delta_\chi(\vp_2),
\end{align}
with transfer functions $t_1$ and $t_2$ calibrated using cross-correlations in simulations as described in \app{TransferFcns}.

We still need to specify how we solve \eqq{DisplIsFilteredNonl} for the Lagrangian-to-Eulerian displacement $\vchi$, or its divergence $\delta_\chi$.
One could solve this equation perturbatively; we will present the relevant formulas in Appendix \ref{app:RecModel}.
Here we will think of $\vchi$ as a `migration' map that shows where each final dark matter particle originated. 
We can hope to estimate this because (a) dark matter particles do not migrate very far over the history of the Universe (typically a few comoving Mpc), (b) they preferentially travel along potential gradients that can be estimated, and (c) they originate from a nearly uniform initial distribution because the clustering power spectrum is small at early times.
This suggests to compile the migration map $\vchi$ iteratively by estimating potential gradients from the observed late-time catalog of galaxies or matter overdensities, moving them back along those gradients, computing potential gradients again from the new catalog, moving them further back, and so on. 
The density of the displaced particles decreases from step to step, until it converges to a set of nearly uniformly distributed particles that will not be moved further because potential gradients nearly vanish.
The total displacement from the late-time nonlinear catalog to the nearly uniform catalog is our estimate of the migration map or Lagrangian-to-Eulerian displacement $\vchi$.

An alternative motivation for this iterative procedure to solve \eqq{DisplIsFilteredNonl} follows from the Eulerian continuity equation
\begin{equation}
  \label{eq:4}
  \delta'+\vnabla\cdot[(1+\delta)\vv]=0.
\end{equation}
This can be written as a convective derivative
\begin{equation}
  \label{eq:5}
  \frac{\mathrm{D}}{\mathrm{D}\tau}\ln(1+\delta) = -\vnabla\cdot\vv=-\vnabla\cdot\frac{\mathrm{d}\vchi}{\mathrm{d}\tau}
= -\frac{\mathrm{d}}{\mathrm{d}\tau}\vnabla\cdot\vc.
\end{equation}
We can solve this equation iteratively in small time steps, where in each step
\begin{equation}
  \label{eq:6}
  \vnabla\cdot\vchi = -\ln(1+\delta_R)\approx -\delta_R.
\end{equation}
Here $\delta_R$ is the nonlinear density filtered with a progressively smaller smoothing scale $R$ that ensures $\delta_R\lesssim 1$ for each step.

Note that the solution of \eqq{DisplIsFilteredNonl} is not unique after shell crossing. So we should not be too concerned with the specific scheme we use to solve the equation. Different choices  will result in  solutions that differ on small scales where they cannot be trusted. 

\section{Numerical setup}
\label{se:NumericalSetup}

Having motivated and described our reconstruction algorithm in the last section, we next describe a concrete numerical implementation of that algorithm and the numerical setup that we will use in \se{Results} to assess the performance of the reconstruction.

\subsection{New reconstruction algorithm}
\label{se:RecRecipe}

\subsubsection{Procedure}
As described above, our reconstruction algorithm consists of two stages. In the first stage, we compute a displacement field $\vchi$ using an iterative scheme. This is an approximation of the true displacement between initial and final conditions.
In the second stage, that displacement is used to estimate the linear density.
In detail, we implement the new reconstruction using the following recipe.

$ $\\
\emph{Stage I: Iteratively solve for displacement field $\vchi$}
\begin{enumerate}
\item Iteratively displace objects as follows:
  \begin{enumerate}
  \item From the catalog, compute the fractional overdensity, $\delta(\vx)=\rho(\vx)/\bar{\rho}-1$, on a regular grid using cloud-in-cell.
  \item Apply Gaussian smoothing with kernel $W_R=e^{-(kR)^2/2}$, with smoothing scale
    \begin{align}
      \label{eq:12}
R=\mathrm{max}(\epsilon_R^{n-1}R_\mathrm{init}, R_\mathrm{min}).
    \end{align}
in the $n$th iteration step,
and truncate small-scale modes by setting $\delta(\vk)=0$ if $k>k_\mathrm{max}$.
  \item Compute the Zeldovich displacement $\vs(\vk)=-\epsilon_s \frac{i\vk}{k^2}\delta(\vk)$ on the grid.
  \item For every object in the catalog, interpolate the displacement $\vs(\vx)$ from the regular grid to the object's position, move the object by that displacement, and update the object's position in the catalog.
  \end{enumerate}
\item For every object in the catalog, compute its total displacement $\vchi(\hat\vq_\mathrm{end})=\hat\vq_\mathrm{end}-\vx_\mathrm{start}$, where $\vx_\mathrm{start}$ is the observed starting position before the first iteration step, and $\hat\vq_\mathrm{end}$ is the end position of the object after iteratively displacing it ($\vx_\mathrm{start}$ is the observed Eulerian position, and $\hat\vq_\mathrm{end}$ is the estimated associated Lagrangian position).
\item Paint\footnote{At every grid point, we average the displacement $\vchi(\hat\vq_\mathrm{end})$ over all nearby objects using the cloud-in-cell distance weight.
If a grid cell is empty, which can happen if the grid is finer than the typical particle separation, we fill $\vchi$ with a random neighbor cell until all cells are filled.
This avoids a broadband power deficit that would result from wrongly setting $\vchi=0$ in empty cells, but it does not affect the correlation coefficient or BAO much.} 
the displacement $\vchi(\hat\vq_\mathrm{end})$, defined at the estimated Lagrangian positions $\hat\vq_\mathrm{end}$, to a regular grid,
and truncate small-scale modes by setting $\vchi(\vk)=0$ if $k>k_\mathrm{max}$.
\end{enumerate}
$ $\\
\emph{Stage II: Given $\vchi$, estimate linear initial density}
\begin{itemize}
\item $\mathcal{O}(1)$ \emph{reconstruction:} 
Compute $\hat\delta_0^{[1]}=\delta_\chi=\vnabla\cdot\vchi$.
\item \emph{$\mathcal{O}(2)$ reconstruction}: 
Using the transfer functions and definitions in \app{TransferFcns}, compute
 \begin{align}
        \label{eq:8}
\quad\hat\delta_\mathrm{0}^{[2]}(\vk)=t_1(k)\delta_\chi(\vk) + t_2(k)\delta_\chi^{[2]}(\vk).
      \end{align}
\end{itemize}

We apply this procedure to dark matter (DM) particles in simulations in real space, leaving important extensions like galaxy bias and redshift space distortions for future work.

\subsubsection{Implementation}
The algorithm involves only simple operations such as moving objects by a displacement field or computing the divergence of the displacement, which are the same basic operations also used by standard reconstruction.
Implementing our method is therefore not more difficult than standard reconstruction \cite{EisensteinRec}.
The computational cost is rather modest, only a few times higher than that of standard reconstruction if running with our default of eight iteration steps.
We have implemented the reconstruction algorithm in a simple Python code. 
With this, reconstructing initial conditions from a catalog of $85\times 10^6$ particles using a $512^3$ grid takes only $3$ CPU hours, which is negligible compared to other computations that are required to analyze LSS data.
The low computational cost and simple operational structure allow the algorithm to be applied to large catalogs, requiring only minimal modifications of codes that already implement the standard reconstruction of \cite{EisensteinRec}.

For application to real data from galaxy redshift surveys, the algorithm would have to be adopted to deal, for example, with the survey selection function including possible gaps in the data, finite shot noise, and redshift space distortions.  
While we do not address these issues here, we note that \cite{Padmanabhan2012BAORec,2014MNRAS.441.3524K} developed approaches to apply standard reconstruction to real data, and those should in principle also be applicable to our method.  
For example, the Zeldovich displacement in each iteration step could be computed by numerically solving the differential equation relating the displacement field and the density in configuration space rather than working on a grid and solving the equation in Fourier space.

\subsubsection{Parameters}
\label{se:recparams}

Our reconstruction algorithm has several parameters.
Throughout the paper we use the following \emph{default parameters} unless explicitly stated otherwise.
We start the iteration with the smoothing scale $R_\mathrm{init}=10\hMpc$.
We then run $N_\mathrm{steps}=8$ iteration steps and halve the smoothing scale from one step to the next, $\epsilon_R=0.5$, until we reach the minimum smoothing scale $R_\mathrm{min}=1.01\,L/N_\mathrm{grid}$.  
In each iteration step we displace particles by the full Zeldovich displacement, $\epsilon_s=1$.
To represent fields on a regular grid, we choose $N_\mathrm{grid}=512$ and truncate modes with $k>k_\mathrm{max}=2\pi/L\times N_\mathrm{grid}/2$.

We discuss these choices in \app{recparamsAppendix}, noting that results are quite robust as long as the smoothing scale is progressively reduced from step to step, $\epsilon_R\sim 0.5$, and the initial smoothing scale is chosen relatively large, $R_\mathrm{init}\gtrsim 5\hMpc$, to ensure $W_R\delta\lesssim 1$.

The $\mathcal{O}(2)$ reconstruction requires two transfer functions $t_1(k)$ and $t_2(k)$ that are smooth functions of the wave vector modulus $k$.
We calibrate them using simulations as described in \app{TransferFcns}.  
The $\mathcal{O}(1)$ reconstruction, in contrast, is just given by $\delta_\chi=\vnabla\cdot\vchi$ and does not involve any transfer function.

For the case of dark matter in real space considered in this paper the reconstruction does not assume any cosmological model.  
This may change if the method is extended to biased tracers in redshift space.

\subsection{Standard reconstruction}
\label{se:StdRecRecipe}
To compare methods, we also run the standard reconstruction method of \cite{EisensteinRec} on all our simulations. 
We implement this by making the following changes to the above recipe: 
Modify the first stage to run only one iteration step with $\epsilon_s=1$ and $R=10\hMpc$;
define $\tilde\vchi(\vx_\mathrm{start})=\hat\vq_\mathrm{end}-\vx_\mathrm{start}$ as a function of the Eulerian starting position before displacing any objects;
paint this $\tilde\vchi$ to a regular grid;
generate a uniform catalog with particles on that regular grid and displace them by $\tilde\vchi$;
call the density of that shifted uniform catalog $\delta_s$;
also compute the density of the particles at $\hat\vq_\mathrm{end}$ after they were displaced, and call the corresponding density $\delta_d$; 
and then $\hat\delta_0^\mathrm{std}=\delta_d-\delta_s$ is the reconstructed density.

A simple extension of this standard method is to use more than one iteration step for the displacement $\vchi$, similar to \cite{SeoBAOShift}.
We discuss this in \app{ExtendedStdRec}.

\subsection{Simulations}
\label{se:sims}

We use simulations produced with the \textsf{FastPM} particle-mesh code \cite{fastpm}.
They closely resemble full $N$-body simulations at much lower computational cost.
To study how far into the nonlinear regime reconstruction works, we run a small-volume simulation with box size $L=500\hMpc$ and $120$ time steps linearly spaced between $a=0.01$ and $a=1$.
To study the BAO signature that is only measurable on relatively large scales, $k\lesssim 0.5\ihMpc$, we use ten large-volume simulations from Ding~\emph{et al.~}\cite{DingEtAl} with $L=1380\hMpc$ and $40$ time steps linearly spaced between $a=0.1$ and $a=1$.
Each of the ten realizations was evolved from initial conditions with and without BAO signature in the initial power spectrum to allow for cosmic variance cancellation when studying the BAO \cite{pacoWiggleNowiggleSims,Marcel1508,DingEtAl,IsoRecWang1703BAO}.
The 'nowiggle' power spectrum without the BAO signature was matched to the broadband shape of the 'wiggle' power spectrum with the BAO signature, using the method described in the appendix of Ref.~\cite{Vlah1509}.
The simulations evolve $2048^3$ particles by computing forces on a $4096^3$ particle-mesh grid.
We take snapshots at redshift $z=0.6$, which is slightly higher than the peak redshift of the SDSS BOSS CMASS galaxy sample (e.g.~\cite{BOSSFinalAlam}), and today at redshift $z=0$.
To speed up computations, we select a $1\%$ subsample that contains roughly $85\times 10^6$ DM particles.
To study the BAO, we subsample the same particles in wiggle and nowiggle simulations by keeping every particle whose particle ID is divisible by 101 (a prime number that avoids undesired regularities in the selection).

The simulations assume a flat $\Lambda$CDM cosmology with $\Omega_m=0.3075, \Omega_ch^2=0.1188$, $\Omega_bh^2=0.0223$, $h=0.6774$ and $\sigma_8=0.8159$ based on \cite{PlanckParams2015}. 
The fiducial BAO scale, given by the radius of the sound horizon at the drag epoch when photons decoupled from baryons, evaluates to $r_\mathrm{BAO}^\mathrm{fid}\approx 147.5\,\mathrm{Mpc}=99.9\hMpc$, assuming the approximation of Eq.~(16) in \cite{Aubourg1411} for simplicity.

\section{Results}
\label{se:Results}

\begin{figure}[tbhp]
\includegraphics[width=0.42\textwidth]{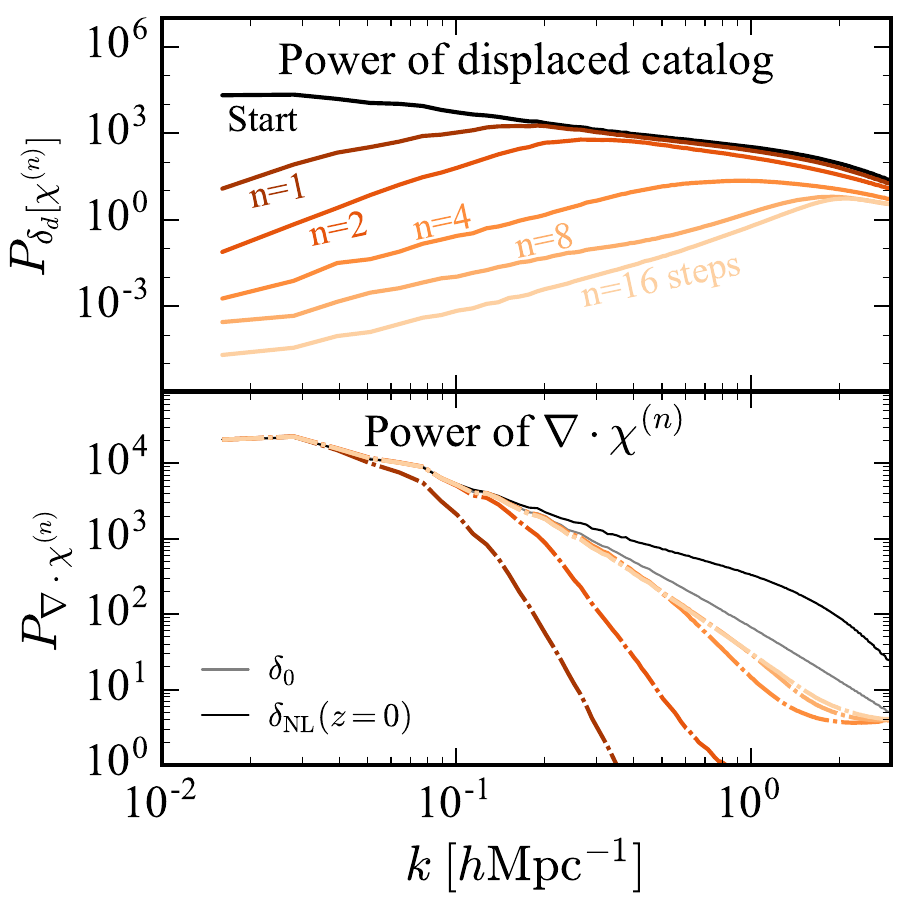}
\caption{The power spectrum of the displaced catalog in the first stage of the reconstruction decreases with the number of iteration steps $n$ (upper panel).  At the same time, the power of the cumulative displacement $\vchi$ increases and approaches the linear power spectrum (lower panel). 
All power spectra are measured from a $L=500\hMpc$ simulation at $z=0$. }
\label{fig:PowerOfDisplacedCat}
\end{figure}

We first check the basic functionality of the iterative procedure in the first reconstruction stage that progressively displaces the DM particles along their potential gradients until a nearly uniform set of particles is obtained.
The upper panel of \fig{PowerOfDisplacedCat} shows the power spectrum of the displaced catalog after $1, 2, 4, 8$, and $16$ steps.
As expected, the power decreases from step to step. After eight steps the large-scale power is reduced by a factor of $10^8$ at $z=0$.
Using more than eight steps does not reduce the large-scale power much further, so that our default choice is to stop after eight iteration steps.
On smaller scales, the power of the displaced catalog is still reduced, but by a much smaller factor, for example a factor of $\sim 100$ at $k=1\ihMpc$.
In this regime, the displacement obtained from the iterative procedure is not unique because of shell crossing as discussed in \se{TheoryMotivation} above.
The lower panel of \fig{PowerOfDisplacedCat} shows how the power of the cumulative displacement $\vchi$ progressively increases with more iterative steps as the power of the displaced catalog decreases.
On large scales, the power of the displacement divergence approaches that of the linear field, but it misses power on small scales where shell crossing happens.
This shows that the basic concepts motivating the reconstruction algorithm in \se{TheoryMotivation} work in the simulations.

To evaluate the numerical performance of our method in more detail we consider several metrics in the next sections, including density slices and histograms, the cross-correlation coefficient with the initial conditions, the BAO signature in the power spectrum, and the full broadband shape of the power spectrum.

\subsection{Qualitative behavior of the reconstructed density}

\begin{figure*}[p]
\includegraphics[width=0.98\textwidth]{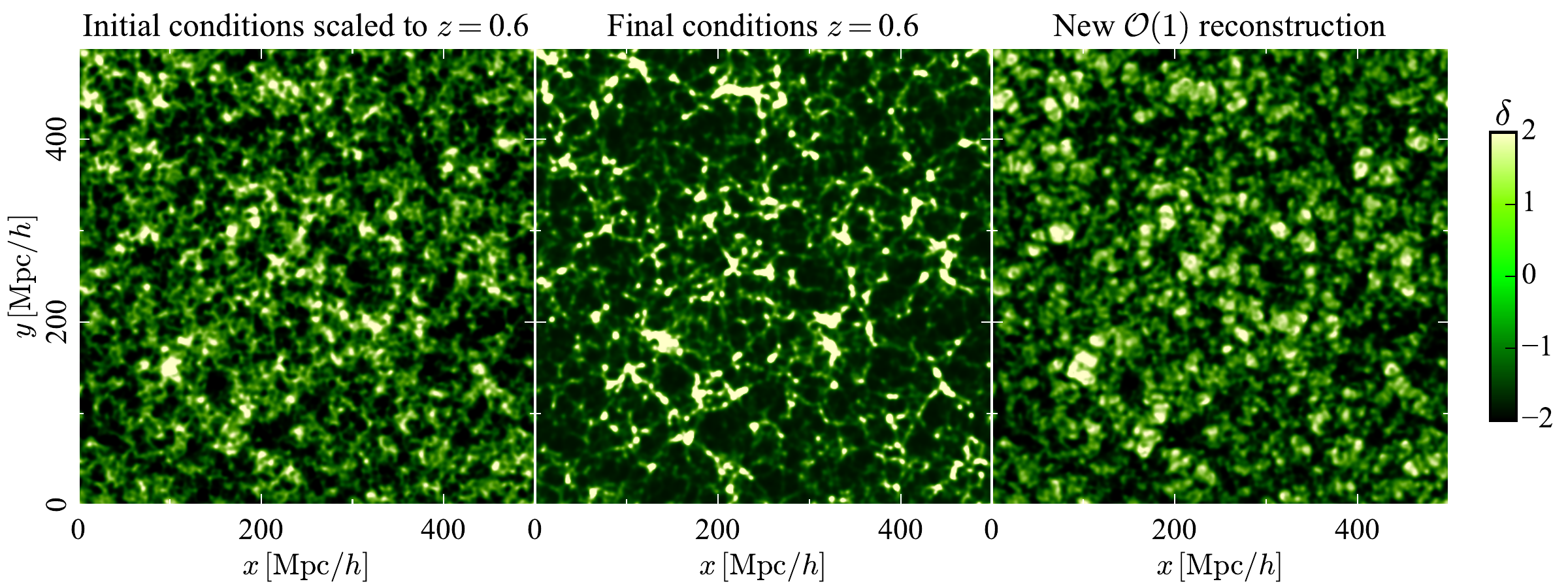}
\caption{Two-dimensional slices of the fractional dark matter overdensity $\delta=\rho/\bar{\rho}-1$ smoothed with a $R=2\hMpc$ Gaussian.
\emph{Left:} Linear initial density linearly scaled to redshift $z=0.6$, showing how the Universe would look if it underwent only linear evolution up to $z=0.6$ and all Fourier modes were Gaussian and linear.
\emph{Middle:} Simulated nonlinear dark matter density at $z=0.6$.
\emph{Right:} Reconstructed density $\delta_\chi=\vnabla\cdot\vchi$, using the default settings described in \se{recparams}.
 }
\label{fig:slices}
\end{figure*}

\begin{figure*}[p]
\includegraphics[width=0.9\textwidth]{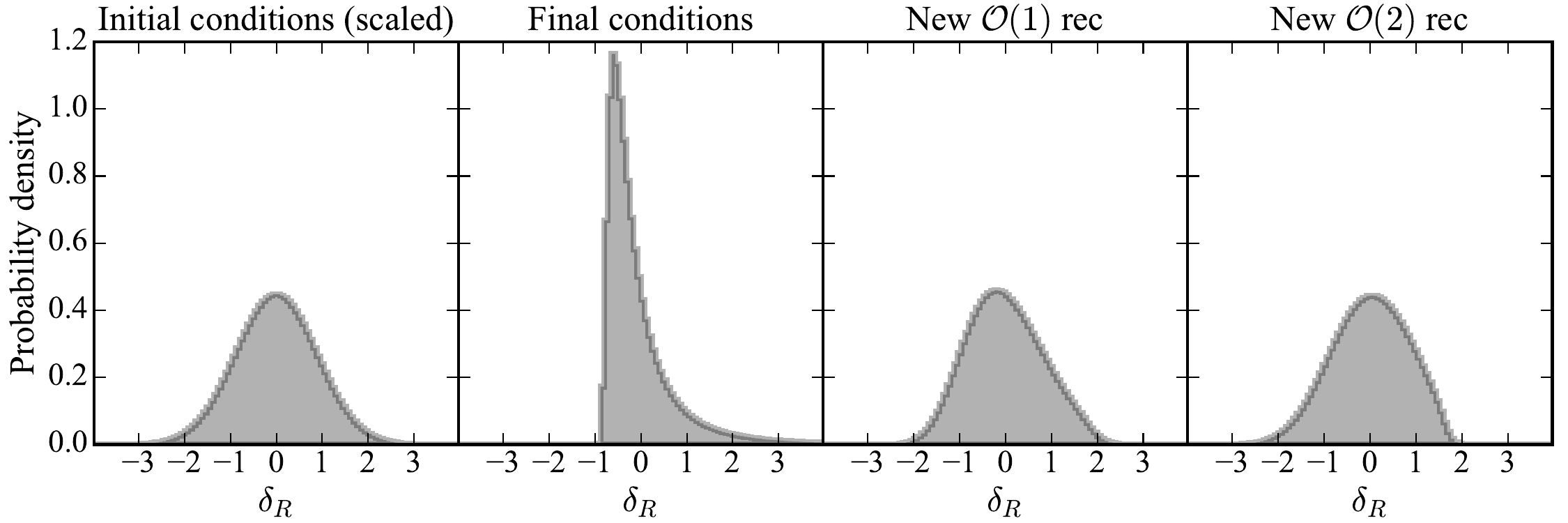}
\caption{Reconstruction makes the density more Gaussian, transferring cosmological information from skewness and higher-order moments to the variance.
From left to right, the panels show histograms of the linear initial condition density linearly scaled to $z=0.6$, the nonlinear DM density at $z=0.6$, the first-order reconstructed density $\vnabla\cdot\vchi$ that involves no transfer functions, and the second-order reconstructed density $\hat\delta_0^{[2]}[\vchi]$ that involves two transfer functions $t_1$ and $t_2$ calibrated to simulations. 
All densities are smoothed with a $R=2\hMpc$ Gaussian filter.
 }
\label{fig:hist}
\end{figure*}

\fig{slices} compares the reconstructed density with the initial conditions in two-dimensional slices through our $L=500\hMpc$ simulation.
The reconstruction successfully recovers the spatial structure of the initial conditions on large scales and in regions with low to moderate overdensities.
On small scales and in regions with high late-time overdensity, however, the reconstruction is less successful at recovering the correct initial conditions because of shell crossing as discussed above.

Besides spatial structure, we can also consider the one-point pdf or histogram of the smoothed mass density as shown in \fig{hist}.
By construction, the linear initial conditions have a Gaussian pdf centered at $\delta=0$.
Under nonlinear gravitational evolution the density turns into a rather non-Gaussian field with a skewed one-point pdf (see \cite{BernardeauReview} for a review).
\fig{hist} demonstrates that reconstruction transforms that nonlinear density back to a field with a one-point pdf that is much closer to a Gaussian again.
In other words, our reconstruction Gaussianizes the mass density and moves cosmological information from the skewness and other higher-order moments back to the variance of the density, similar to findings in \cite{Marcel1508,SlepianBAO3pt,Hikage1703}.

\subsection{Correlation with the linear initial conditions}

The two-dimensional (2D) slices and one-point histograms of the mass density are useful qualitative measures, indicating that the reconstruction algorithm can at least partially recover the initial conditions.
To check this more quantitatively and rigorously we turn to Fourier space.

\begin{figure}[tp]
\includegraphics[width=0.48\textwidth]{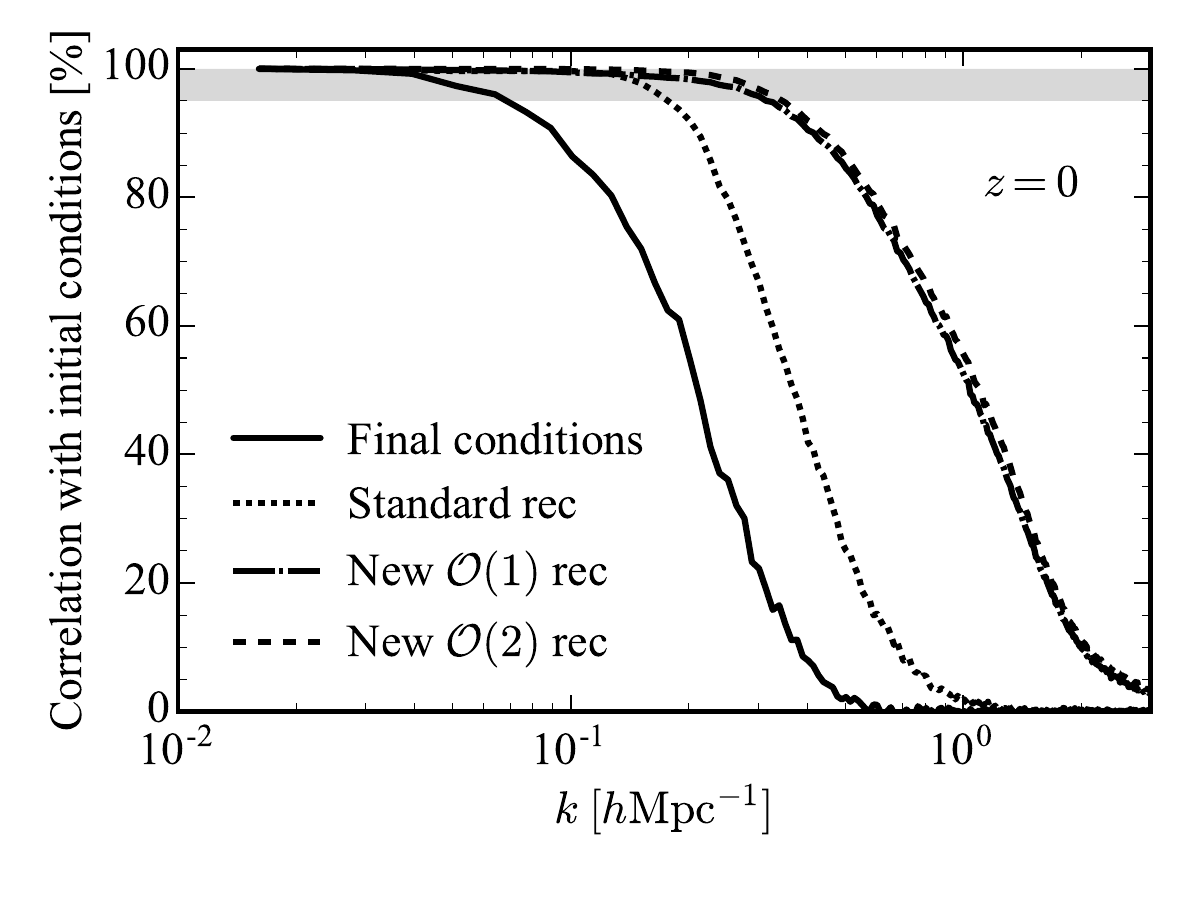}
\includegraphics[width=0.49\textwidth]{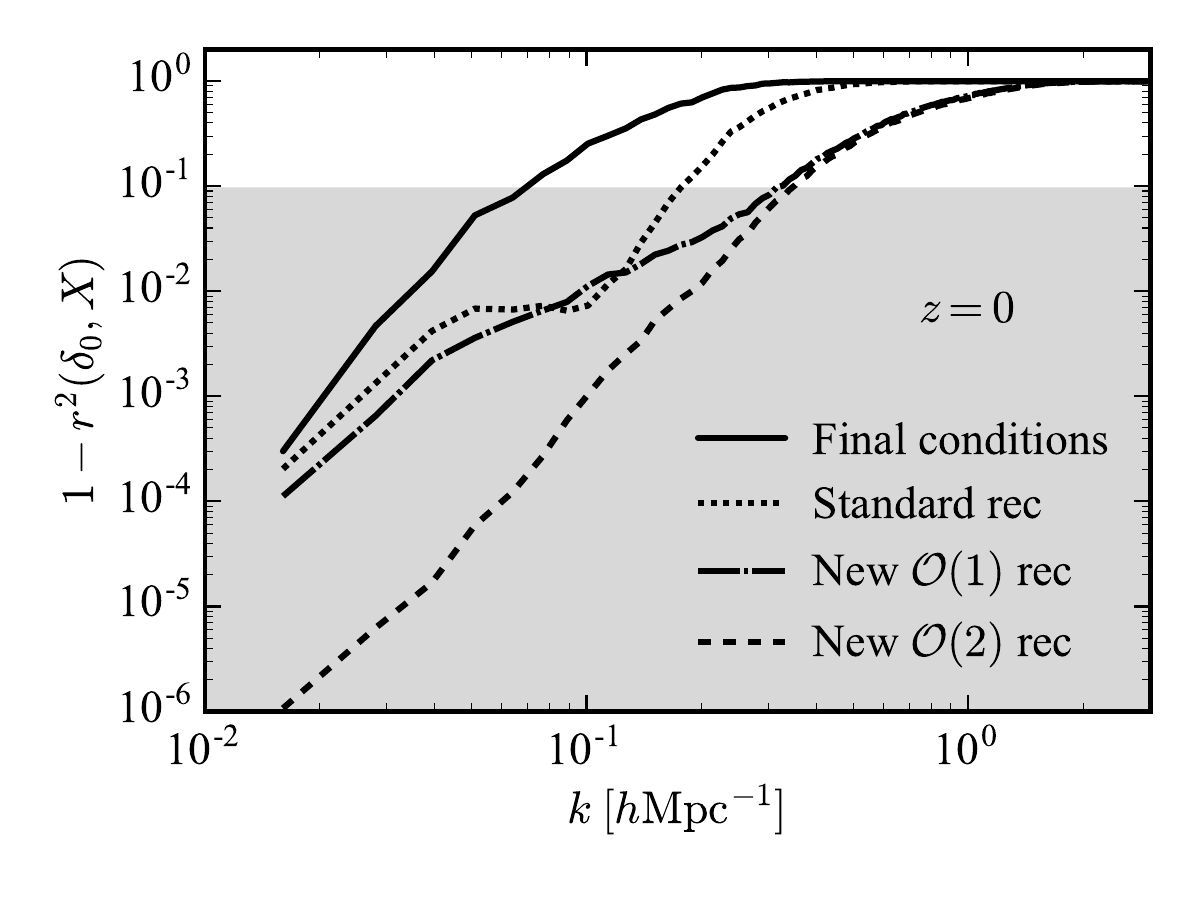}
\caption{\emph{Top:} Correlation coefficient with the linear initial conditions, $r(k)\equiv \la\delta\delta_0\ra/\sqrt{\la\delta\delta\ra\la\delta_0\delta_0\ra}$, as a function of wavenumber $k$.
\emph{Bottom:} Power in the difference between initial and reconstructed fields in units of the power spectrum of the initial field. This is given by one minus the squared correlation coefficient with the initial conditions, $1-r^2(k)$.
In the shaded regions the correlation with the initial conditions is better than $95\%$.
Reconstruction improves the correlation with the initial conditions substantially.
The curves are computed from a $L=500\hMpc$ simulation at redshift $z=0$.
}
\label{fig:rccSimple}
\end{figure}

\fig{rccSimple} shows the cross-correlation coefficient $r(k)$ between the reconstructed density and the linear initial conditions as a function of Fourier wavenumber $k$ measured from our $L=500\hMpc$ simulation at $z=0$.
Using our first-order method, the reconstructed density is more than $95\%$ correlated with the initial conditions on scales $k\le 0.31\ihMpc$.
Using our second-order method, this is the case on scales $k\le 0.35\ihMpc$.
For comparison, the wavenumber where the correlation with the initial conditions drops below $95\%$ is $k=0.18\ihMpc$ for standard reconstruction, and $k=0.07\ihMpc$ for the nonlinear density without reconstruction, in the same simulation.
Based on this correlation coefficient with the linear density, our reconstruction thus improves the $k$ range by a factor of $2$ over standard reconstruction, and by a factor of $5$ compared to performing no reconstruction. 
At redshift $z=0.6$ the improvement factors are similar; see \fig{rccSimplez0pt6} in the appendix.

Related to this, the lower panel of \fig{rccSimple} shows the fractional error of the reconstructed density phases.
This is represented by the power spectrum of the difference between reconstructed density and true linear density, in units of the linear power spectrum.  
This fractional nonlinear error power can be shown to reduce to one minus the squared correlation with the initial conditions \cite{Baldauf:2015zga}.\footnote{Focusing on the density \emph{phase} correlation, we have implicitly assumed here that the reconstructed density is rescaled by $\mathrm{TF}(k)\equiv\langle\hat\delta_0\delta_0\rangle/\langle\delta_0\delta_0\rangle$ as in Eq.~4.6 in \cite{Baldauf:2015zga}.  We will discuss the density \emph{amplitude} later in \se{Pk} and \app{TransferFcns}.}
\fig{rccSimple} demonstrates that our reconstruction significantly reduces this nonlinear error and improves the correlation with the initial conditions on all scales, outperforming the standard method.
For our second-order method, the nonlinear error power relative to the linear power is
$1-r^2\simeq (10^{-6},10^{-4},10^{-3},10^{-2})$ at $k=(0.02,0.06,0.1,0.2)\ihMpc$ at $z=0$.
At higher redshift, $z=0.6$, the nonlinear error at the same scales is slightly smaller, so that $1-r^2\simeq (10^{-6},10^{-4},10^{-3},10^{-2})$ at $k=(0.02,0.1,0.15,0.3)\ihMpc$
as shown in \fig{rccSimplez0pt6} in the appendix.
At both redshifts the nonlinear error is very small, so that for most practical purposes we can regard the reconstructed and initial densities as identical on large scales.

At some point, reconstruction should be limited by the stochastic displacement term identified in \cite{Baldauf:2015tla}, because it prohibits a deterministic mapping between initial and final conditions, at least using perturbation theory.
This would imply that one cannot improve over $1-r^2\simeq (5\times 10^{-5},5\times 10^{-4},10^{-2})$ at $k=(0.06,0.1,0.2)\ihMpc$ at $z=0$, as shown in Fig.~22 of \cite{Baldauf:2015tla}.
Our second-order reconstruction reaches that limit within a factor of about 2.
We therefore expect that other reconstruction methods could improve over our method by at most a factor of 2 on large scales.

As motivated in the Introduction, using our method to recover initial conditions over a wide range of scales can substantially improve many of the science goals of galaxy surveys by increasing the number of linear Fourier modes amenable to cosmological analysis. 
An important caveat is, however, that our numerical setup is rather idealized because we work with DM particles and ignore galaxy biasing and redshift space distortions. 
Both effects will certainly degrade the performance of reconstruction in practical applications.
We plan to study this in future work.

As described in \se{NumericalSetup}, we use the eight-step displacement $\vchi^{(8)}$ for our new reconstruction, but the one-step displacement $\vchi^{(1)}$ for the standard reconstruction, because this is what has been used in the literature so far. 
\fig{1mr2Steps} in \app{HighZ} explores how the performance depends on the number of steps used to construct the displacement field $\vchi$.
We see only little benefit in using more than eight iteration steps, indicating that the algorithm has converged after eight steps.
Extending the standard reconstruction by applying it to the eight-step displacement $\vchi^{(8)}$ improves over using $\vchi^{(1)}$ on most scales, but it still performs worse than our second-order reconstruction (see \app{ExtendedStdRec} for discussion).

For the correlation coefficient shown in \fig{rccSimple}, the first order reconstruction does not depend on transfer functions because any rescaling by a function of $k$ would not affect the correlation coefficient.
In contrast, the second order correction does require transfer functions that were calibrated to simulations as described in \app{TransferFcns}. Just as in the case of forward modeling in \cite{Baldauf:2015tla}, the shape of the transfer functions can probably be understood using the EFT approach, but we leave this for future work.

\subsection{Baryonic acoustic oscillations}

\begin{figure}[htp]
\includegraphics[width=0.48\textwidth]{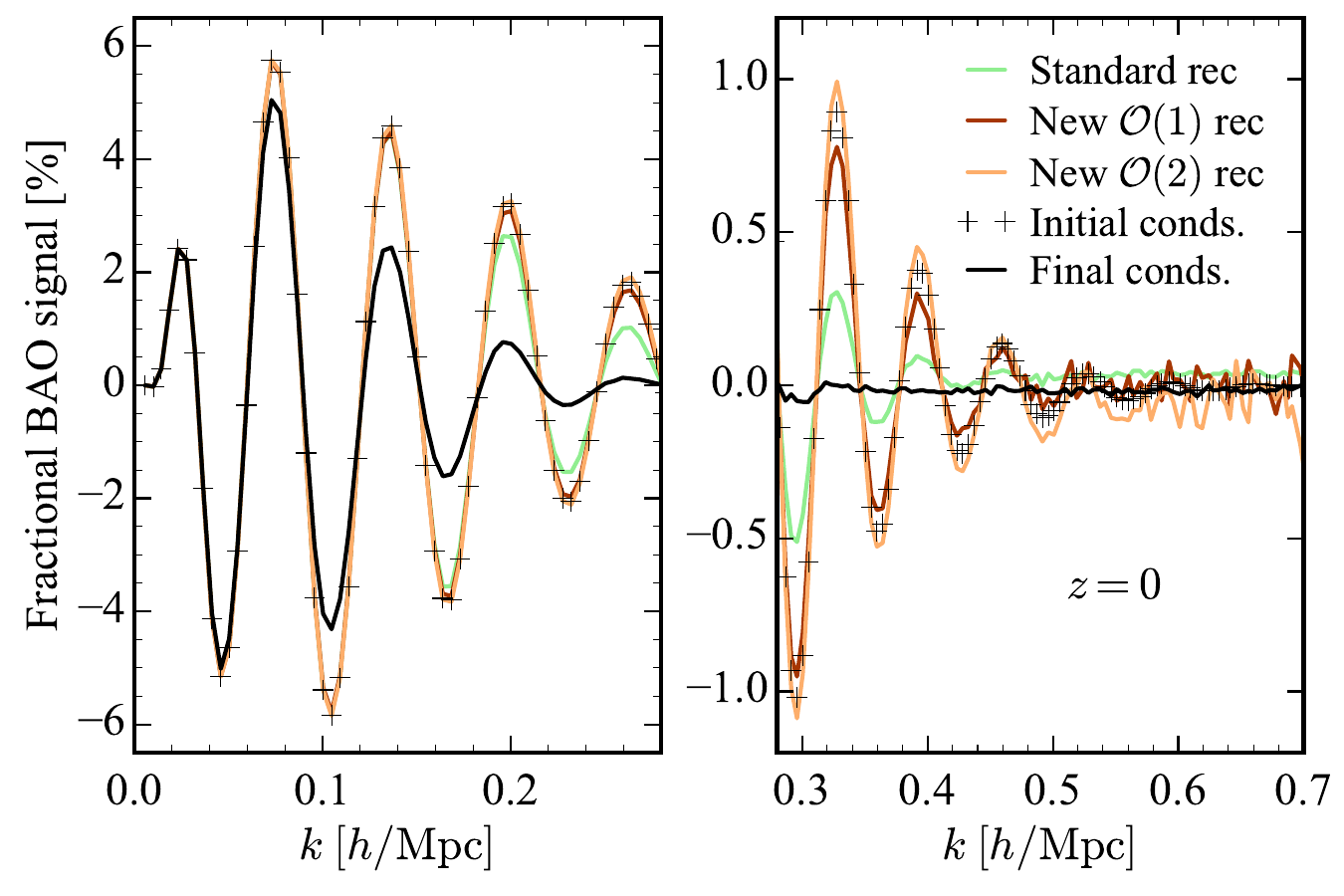}
\caption{Fractional BAO signal in the power spectrum, given by the fractional difference of simulations initialized with and without BAO wiggles,
$\langle(\hat P_\mathrm{wiggle}-\hat P_\mathrm{nowiggle})\rangle/\langle\hat P_\mathrm{nowiggle}\rangle$.
Reconstruction sharpens the BAO wiggles so that they agree with those in the linear initial conditions.
The power spectra are averaged over ten large-volume simulations at $z=0$,
using the same random seed for each wiggle and nowiggle simulation to cancel most of the cosmic variance \cite{pacoWiggleNowiggleSims,Marcel1508,DingEtAl,IsoRecWang1703BAO}.
}
\label{fig:BAOsignal}
\end{figure}

Measurements of the BAO scale from the galaxy power spectrum are a prime example for the application of reconstruction, because it reverses or avoids large-scale shifts that would otherwise wash out the BAO wiggles in the observed galaxy power spectrum, degrading the measurement \cite{EisensteinRec,EisensteinSeoWhite0604,Padmanabhan0812}. 
As mentioned above, the standard reconstruction technique has been successfully applied to several redshift surveys, improving the precision of the measured BAO scale typically by a factor of $\sim 2$ \cite{Padmanabhan2012BAORec,AndersonBAODR9_1203,1409.3242,Anderson1312.4877,2014MNRAS.441.3524K,Florian1506.03900,BOSSFinalAlam}, with similar improvements expected for future surveys.
It is therefore exciting to see if our method can improve BAO measurements further.
To answer this, we use ten large-volume simulations with $L=1380\hMpc$ that were produced by Ding~\emph{et al.}~\cite{DingEtAl}  as described in \se{sims} above.

\fig{BAOsignal} shows the fractional BAO \emph{signal} in the simulations.
Our method restores the BAO signal of the linear density perfectly, reversing the nonlinear damping.
This is not surprising given that the BAO signal is only visible at $k<0.5\ihMpc$, where the reconstructed density is more than $90\%$ correlated with the linear density as we already found in \fig{rccSimple}.
Standard reconstruction  (green line in \fig{BAOsignal}) also reduces the nonlinear damping, but it does not recover the full linear BAO wiggles at $k\gtrsim 0.2\ihMpc$.

To see if the signal-to-noise ratio of the BAO scale estimated from the power spectrum is also improved by reconstruction, we need to characterize the \emph{noise} of the estimated BAO scale. 
This would be straightforward if we knew the covariance between power spectrum bins after reconstruction, but that is difficult to compute reliably.
We therefore choose a simpler Monte Carlo approach and estimate the BAO uncertainty from the scatter of the best-fit BAO scale among the ten simulations. 
This provides a conservative estimate for the uncertainty of the best-fit BAO scale (see \app{BAOFitting}, where we also describe our fitting procedure).

\begin{figure*}[tp]
\includegraphics[width=0.75\textwidth]{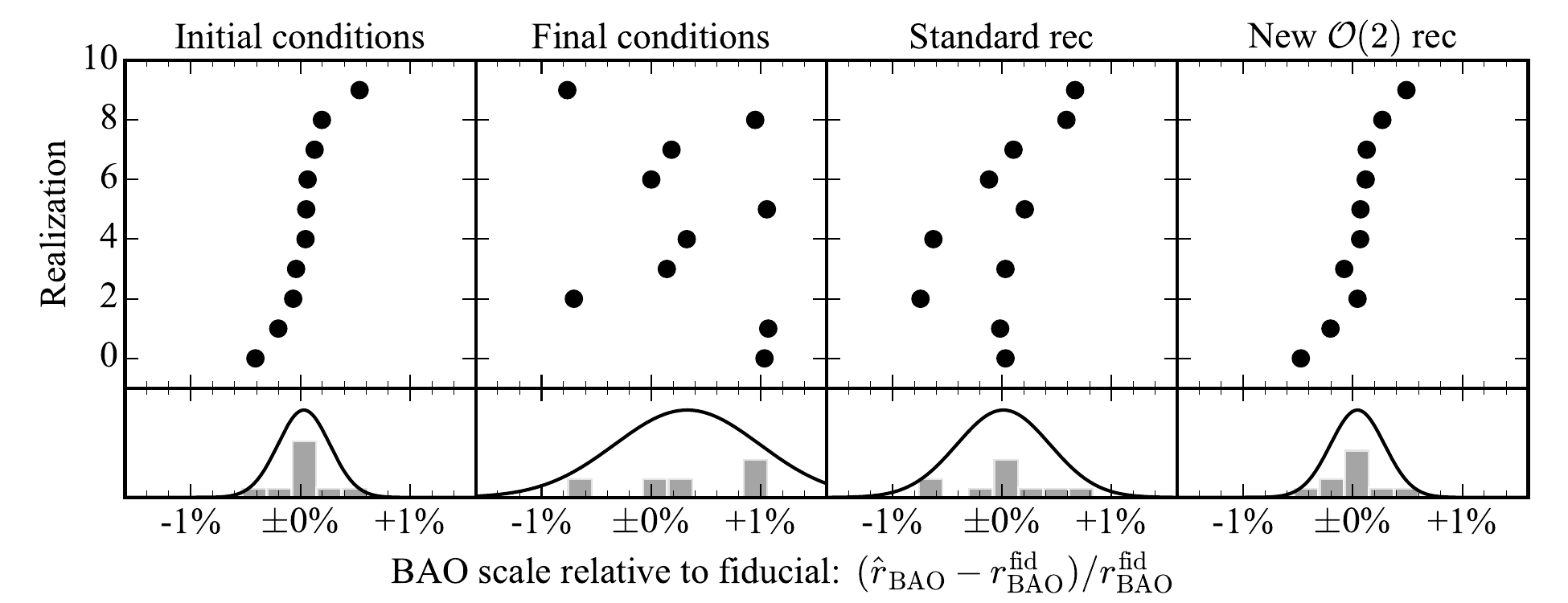}
\caption{Fractional bias of the best-fit BAO scale relative to the fiducial BAO scale in ten $2.6\,h^{-3}\mathrm{Gpc}^3$ simulations at $z=0$.
In each simulation, the BAO scale is estimated by fitting a model to the measured power spectrum at $k\le 0.6\ihMpc$ as described in \app{BAOFitting}.
The lower subpanels show histograms of the best-fit BAO scale (grey), and corresponding Gaussian pdfs (solid black) based on sample mean and sample standard deviation of the best-fit BAO scale.
The realizations are sorted according to their initial linear BAO scale.
 }
\label{fig:histalpha}
\end{figure*}

\begin{figure}[tp]
\includegraphics[width=0.48\textwidth]{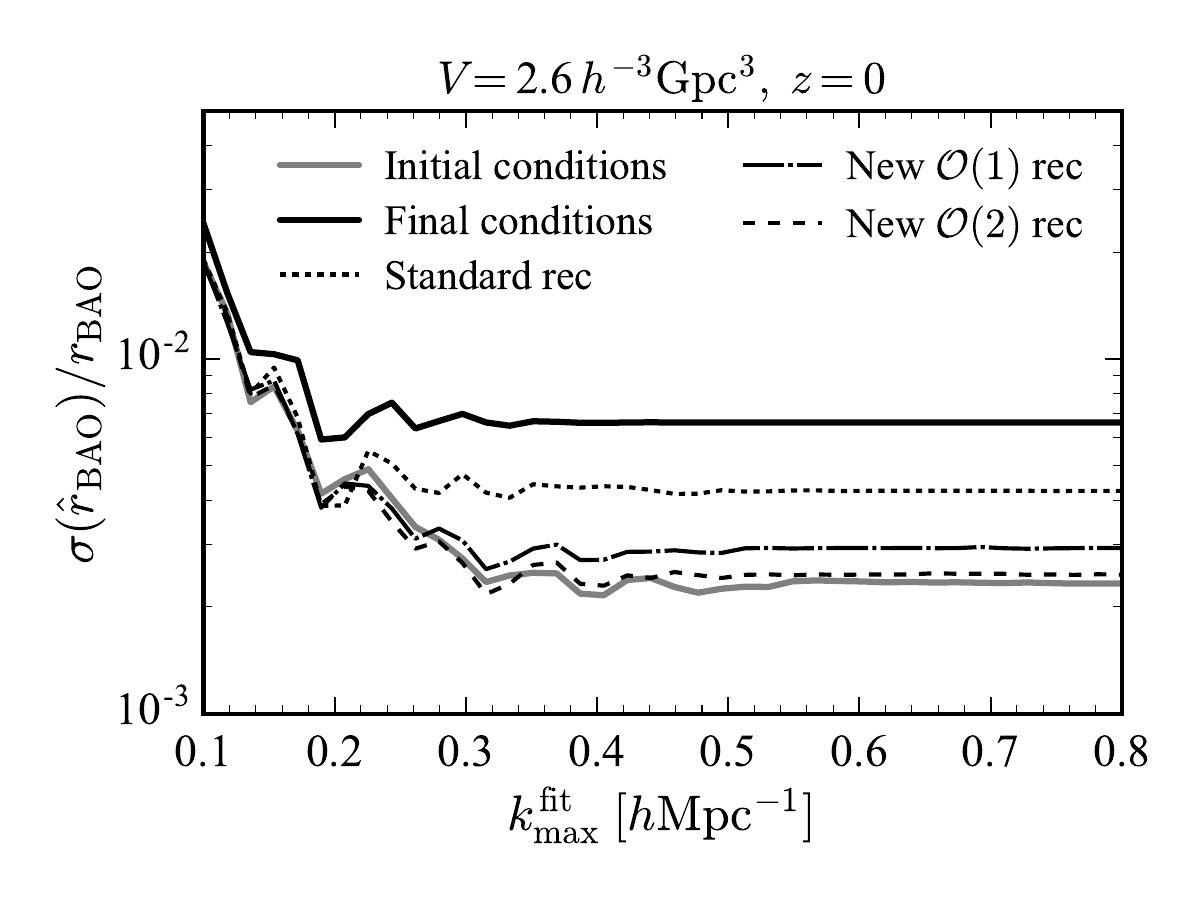}
\caption{Fractional BAO error bar as a function of maximum wavenumber used for fitting the BAO scale.
The error bar is a Monte Carlo estimate obtained from ten simulations at $z=0$ with $V=2.6\,h^{-3}\mathrm{Gpc}^3$ each:
We fit the BAO scale to the ratio of wiggle and nowiggle power spectra in each of the ten simulations, and then compute the scatter of the best-fit BAO scale across the ten simulations.
The iterative $\mathcal{O}(2)$ reconstruction matches the linear initial conditions perfectly.
}
\label{fig:BAOkmax}
\end{figure}

\begin{table}[tp]
\centering
\renewcommand{\arraystretch}{1.0}
\begin{tabular}{@{}p{2.0cm}lllll@{}}
\toprule
& \phantom{} & \multicolumn{3}{l}{Mean BAO scale}  \\
Field && vs lin.~theory &\phantom{$\,$}& vs lin.~realization \\
\colrule
Initial conds. && $+0.05\,\mathrm{Mpc}$ [$+0.03\%$] && $+0.00\,\mathrm{Mpc}$ [$\pm0.00\%$] \\
Final conds. && $+0.49\,\mathrm{Mpc}$ [$+0.33\%$] && $+0.44\,\mathrm{Mpc}$ [$\mathbf{+0.30}\%$] \\
Standard rec && $+0.02\,\mathrm{Mpc}$ [$+0.01\%$] && $-0.03\,\mathrm{Mpc}$ [$-0.02\%$] \\
New $\mathcal{O}(1)$ rec && $+0.05\,\mathrm{Mpc}$ [$+0.03\%$] && $\pm 0.00\,\mathrm{Mpc}$ [$\pm 0.00\%$] \\
New $\mathcal{O}(2)$ rec && $+0.06\,\mathrm{Mpc}$ [$+0.04\%$] && $+0.02\,\mathrm{Mpc}$ [$\mathbf{+0.01}\%$] \\
\botrule
\end{tabular}
\caption{Systematic bias of the BAO scale estimated from the best-fit BAO scale from the power spectrum of ten large-volume simulations at $z=0$.
The BAO scale from the nonlinear density is biased high by $0.3\%$.
Reconstruction eliminates that bias \cite{Padmanabhan0812,NohPadmanabhanWhite2009,SherwinZaldarriaga2012,TassevRec,SeoBAOShift,white1504.03677,Mehta1104_RecBias}.
The residual biases after reconstruction are small and likely consistent with zero because the estimates are derived from only ten simulations.
The left column shows the sample mean of the best-fit BAO scale relative to the fiducial theoretical value, $\langle\hat r_\mathrm{BAO}\rangle-r_\mathrm{BAO}^\mathrm{fid}$;
the right column is relative to the initial condition of each simulation,  $\langle \hat r_\mathrm{BAO}-\hat r_\mathrm{BAO}^\mathrm{lin}\rangle$, canceling cosmic variance.
}
\label{tab:BAOFitsMean}
\end{table}

\begin{table}[tp]
\centering
\renewcommand{\arraystretch}{1.0}
\begin{tabular}{@{}p{2.1cm}lllll@{}}
\toprule
& \phantom{} & \multicolumn{3}{l}{Rms scatter of BAO scale} \\
Field && vs lin.~theory &\phantom{$\quad$}& vs lin.~realization \\
\colrule
Initial conds. && $0.35\,\mathrm{Mpc}$ [$0.24\%$] && $0\,\mathrm{Mpc}$ [$0\%$] \\
Final conds. && $0.99\,\mathrm{Mpc}$ [$\mathbf{0.66}\%$] && $1.20\,\mathrm{Mpc}$ [$0.81\%$] \\
Standard rec && $0.63\,\mathrm{Mpc}$ [$0.42\%$] && $0.55\,\mathrm{Mpc}$ [$0.37\%$] \\
New $\mathcal{O}(1)$ rec && $0.44\,\mathrm{Mpc}$ [$0.29\%$] && $0.13\,\mathrm{Mpc}$ [$0.08\%$] \\
New $\mathcal{O}(2)$ rec && $0.37\,\mathrm{Mpc}$ [$\mathbf{0.25}\%$] && $0.08\,\mathrm{Mpc}$ [$0.05\%$] \\
\botrule
\end{tabular}
\caption{\emph{Left column:} Root-mean-square scatter of the best-fit BAO scale between ten $2.6\,h^{-3}\mathrm{Gpc}^3$ simulations at $z=0$.
This is a Monte Carlo estimate for the expected statistical $1\sigma$ uncertainty when measuring the BAO scale from the power spectrum in a single $2.6\,h^{-3}\mathrm{Gpc}^3$ volume.
\emph{Right column:} Rms scatter of the BAO scale relative to that in the initial conditions of each simulation, $\hat r_\mathrm{BAO}-\hat r_\mathrm{BAO}^\mathrm{lin}$, which is sourced by nonlinear shift terms as discussed in \se{ResidualShifts}.
All numbers are somewhat uncertain because they were estimated from the scatter of only ten simulations. 
}
\label{tab:BAOFitsScatter}
\end{table}

\fig{histalpha} compares the best-fit BAO scales estimated from linear initial conditions, nonlinear late-time density, and reconstructed density in each of the ten simulations, by fitting the BAO scale to the power spectrum at $k\le k_\mathrm{max}=0.6\ihMpc$.
This shows that our reconstruction recovers the linear BAO scale with high precision and on a realization-by-realization basis.

To estimate if the estimated BAO scale is systematically biased relative to the true BAO scale, we compute the expectation value of the best-fit BAO scale; see Table~\ref{tab:BAOFitsMean}.
Within the uncertainty of our ten simulations, we do not find evidence for any systematic BAO bias after any of the reconstruction methods that we tested.  
The reconstructions thus eliminate the systematic nonlinear BAO bias of $\sim 0.3\%$ at $z=0$ that is generated by shifts of particles that were separated by the pristine BAO scale in the initial conditions \cite{EisensteinSeoWhite0604,EisensteinRec}, and that would be present when measuring the BAO scale from the nonlinear power spectrum without reconstruction. 
This is consistent with previous findings \cite{SeoBAOShift,Padmanabhan0812,NohPadmanabhanWhite2009,SherwinZaldarriaga2012,TassevRec,white1504.03677,Mehta1104_RecBias}.

To estimate the statistical $1\sigma$ uncertainty corresponding to measuring the BAO scale from the power spectrum in a $2.6\,h^{-3}\mathrm{Gpc}^3$ volume, we compute the root-mean-square (rms) scatter of the best-fit BAO scale between the ten simulations; see Table \ref{tab:BAOFitsScatter} and \fig{BAOkmax}.

The uncertainty of the BAO scale from the nonlinear power spectrum is increased by a factor of 2.8 at $z=0$ and by a factor of 2.6 at $z=0.6$ relative to the uncertainty from the linear power spectrum.
This is again caused by shifts of particles that were separated by the BAO scale in the early universe.
By reducing those shifts, standard reconstruction \cite{EisensteinRec} reduces the statistical BAO uncertainty by a factor of 1.6 at $z=0$ and by a factor of 1.9 at $z=0.6$ relative to performing no reconstruction. 
Standard reconstruction thus reduces the nonlinear degradation of the BAO signature, but it does not reverse it entirely.
In contrast, our iterative second-order reconstruction reduces the BAO uncertainty by a factor of 2.7 at $z=0$ and by a factor of 2.5 at $z=0.6$ relative to performing no reconstruction, i.e.~the nonlinear degradation is fully reversed within the uncertainty of our estimates.
Relative to the standard reconstruction, the improvement from our method ranges from $70\%$ at $z=0$ to $30\%$ at $z=0.6$.

At both redshifts, $z=0.6$ and $z=0$, the statistical BAO uncertainty after the iterative second order reconstruction matches that of the linear initial conditions ($\simeq 0.24\%$), so that our reconstruction cannot be further improved for measuring BAO in an observed volume of $2.6\,h^{-3}\mathrm{Gpc}^3$. We will return to this point in the next section.
Again, an important caveat is that practical issues like galaxy bias, shot noise, and redshift space distortions are expected to degrade the reconstruction performance in practice.

\subsection{Impact of residual nonlinear shifts on BAO}
\label{se:ResidualShifts}

The measured properties of the best-fit BAO scale in the last section have a non-negligible uncertainty because they were obtained from a somewhat small number of only ten realizations.
To provide a more stringent test, we cancel the linear cosmic variance contribution to the BAO uncertainty by measuring the difference of the best-fit BAO scale between initial and reconstructed density in each individual realization.

To motivate this procedure further, it is useful to split the statistical uncertainty of the measured BAO scale into a linear cosmic variance contribution $\sigma_\mathrm{lin}$ that would be present even if we were given the linear density (due to the finite observed volume and finite width of the linear acoustic peak) and an additional nonlinear noise term $\sigma_\mathrm{NL}$ determined by the rms of nonlinear shifts broadening the acoustic peak,
  \begin{align}
    \label{eq:13}
    \sigma^2_\mathrm{survey} = \frac{V_\mathrm{sim}}{V_\mathrm{survey}}\left(\sigma^2_\mathrm{lin}+\sigma^2_\mathrm{NL}\right).
  \end{align}
We included a prefactor to indicate the expected volume scaling.
Reconstruction reduces $\sigma_\mathrm{NL}$ by reducing nonlinear shifts.
At some point the residual shift terms are small enough that their broadening does not matter compared to the intrinsic width of the acoustic peak in the linear density.
Once we reach that threshold, reconstruction cannot improve BAO measurements any further \cite{EisensteinRec}.
In other words, the BAO signal-to-noise ratio can never be better than if we observed the linear density in the same volume.

After our reconstruction, we cannot distinguish the total BAO uncertainty from the linear uncertainty, i.e.~the nonlinear noise term $\sigma_\mathrm{NL}$ is reduced so much that it is not detectable in our simulations.
Our reconstruction thus reduces the rms of nonlinear shifts well below the threshold corresponding to the intrinsic width of the linear acoustic peak.

\begin{figure*}[tp]
\includegraphics[width=0.75\textwidth]{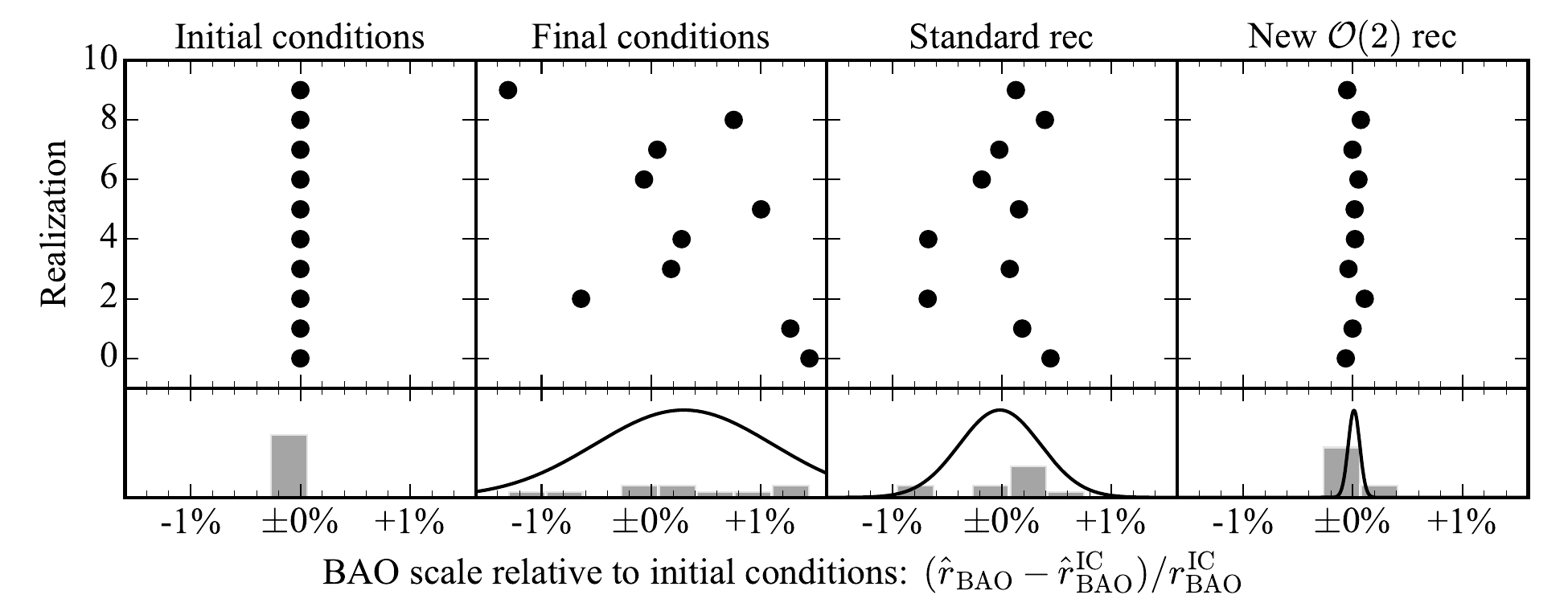}
\caption{Same as \fig{histalpha}, but for each simulation we estimate the BAO scale in the linear initial conditions and subtract it off.
This cancels the cosmic variance caused by linear finite-volume fluctuations of the initial conditions, allowing for a more accurate comparison of methods.
The remaining scatter between simulations corresponds to the BAO uncertainty caused by nonlinear shift terms (see text for discussion).
Our reconstruction reduces this substantially and recovers the linear BAO scale in each individual simulation with high precision.
}

\label{fig:histalphaMinusLinear}
\end{figure*}

\begin{figure}[htp]
\includegraphics[width=0.48\textwidth]{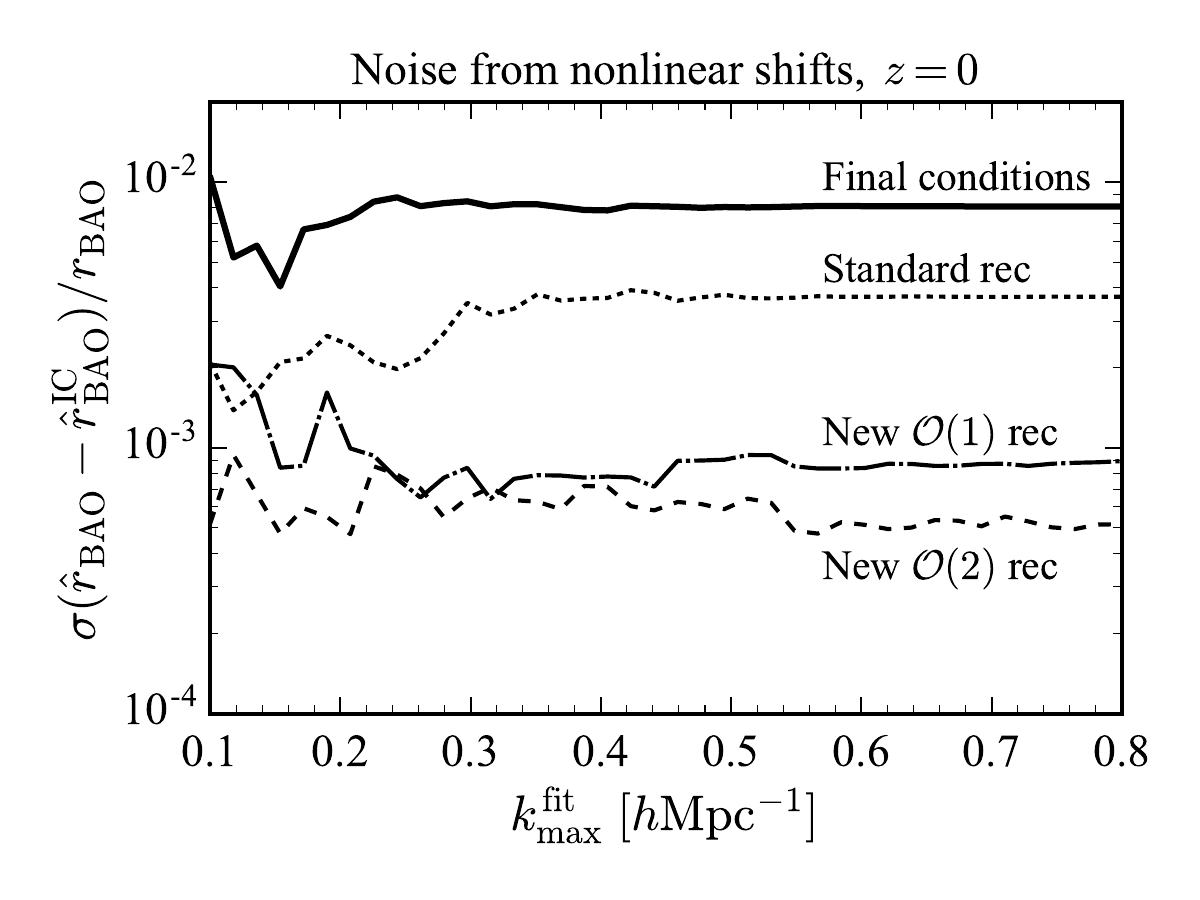}
\caption{Nonlinear BAO noise contribution $\sigma_\mathrm{NL}$ sourced by nonlinear shift terms that wash out the acoustic peak.
The plot shows an estimate of this nonlinear noise divided by the BAO signal, as a function of the maximum wavenumber used to fit for the BAO scale in the power spectrum.
To cancel the linear noise contribution, the linear BAO scale of each simulation is substracted from the measured late-time BAO scale as in \fig{histalphaMinusLinear}; the rms scatter of that difference between simulations is given by nonlinear terms that are not present in the initial conditions.
By construction, the linear density has zero nonlinear noise and is therefore not shown.
Reconstruction reduces the nonlinear noise due to nonlinear shifts significantly. See \se{ResidualShifts} for discussion.
}
\label{fig:BAOkmaxNoCV}
\end{figure}

To see how well different reconstruction methods suppress the nonlinear shift noise $\sigma_\mathrm{NL}$ 
we cancel the linear cosmic variance noise term by subtracting the linear BAO scale of the initial conditions of each simulation, as illustrated in \fig{histalphaMinusLinear}.
This again shows that our reconstruction recovers the correct BAO scale of the initial conditions realization by realization, in a much better way than standard reconstruction.
An estimate for the ratio of nonlinear noise relative to the BAO signal, $\sigma_\mathrm{NL}/r_\mathrm{BAO}$, is shown in \fig{BAOkmaxNoCV} (also see right column of Table~\ref{tab:BAOFitsScatter} for reference).
This demonstrates that the nonlinear noise caused by nonlinear shift terms is substantially reduced by reconstruction.
Relative to the nonlinear density, standard reconstruction reduces the nonlinear noise by a factor of 2.2, our first-order method reduces it by a factor of 9, and our second-order method reduces it by a factor of 15.
We obtain similar factors at redshift $z=0.6$.

Our reconstruction method thus reduces the rms dispersion of nonlinear shifts much more effectively than standard reconstruction.
However, as alluded to above, most of that improvement is not relevant in practice because the acoustic peak of the linear density has some intrinsic width so that reducing the rms of the nonlinear shifts below that width does not help to improve the BAO signal-to-noise ratio further. 
Our reconstruction reduces the nonlinear shifts well below that threshold, easily obtaining the BAO signal-to-noise ratio of the linear density.

The fact that our reconstruction works so much better than it needs to makes us optimistic that it can also restore the linear acoustic peak in more realistic scenarios that account for galaxy bias, redshift space distortions, or higher shot noise.
The efficient suppression of nonlinear shift terms also explains the excellent correlation of the reconstructed density with the linear density. This is useful for applications beyond BAO measurements, for example to extract cosmological information from the broadband power spectrum as we will discuss in the next section, or information about the statistics of primordial fluctuations from higher-order statistics of the reconstructed density.

In summary, the above results show that our iterative reconstruction method is very powerful at reconstructing the BAO scale, and substantially improves over the standard reconstruction method.  
The reconstructed BAO scale is so close to the true BAO scale of the initial conditions that the two cannot be told apart within cosmic variance of current and near-future survey volumes, at least according to our idealized simulations.

\subsection{Full shape of the reconstructed power spectrum}
\label{se:Pk}

\begin{figure}[tbp]
\includegraphics[width=0.47\textwidth]{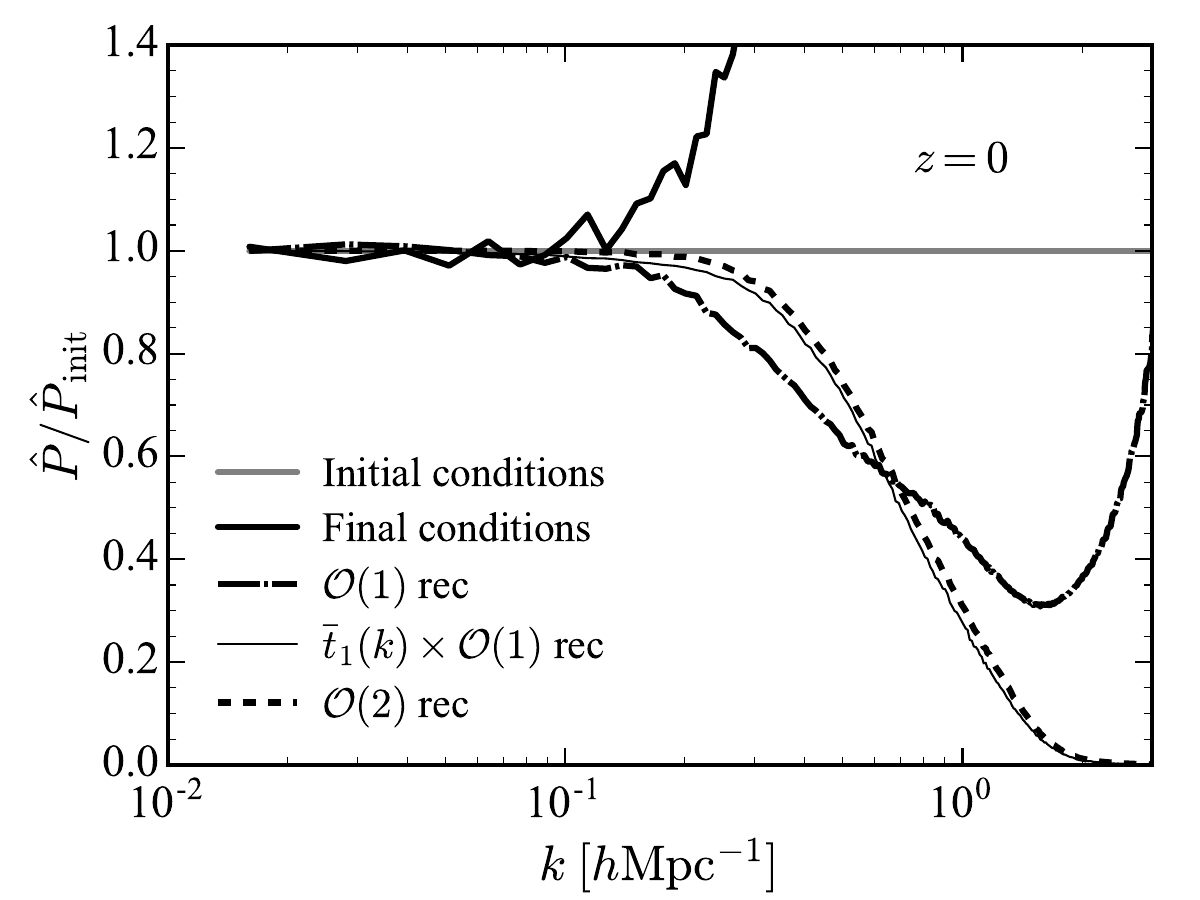}
\caption{Power spectra in our $L=500\hMpc$ simulation at $z=0$, divided by the linear initial power spectrum linearly scaled to $z=0$.
Compared to the nonlinear density without reconstruction (thick solid line), reconstruction significantly improves the agreement with the linear power spectrum on intermediate scales.
Our first-order reconstruction, $\vnabla\cdot\vchi$, has no transfer functions, while the second-order method uses transfer functions discussed in \app{TransferFcns}. 
The spectra are raw spectra without mitigating CIC kernel or shot noise, both of which matter at $k\gtrsim 1\ihMpc$.
The high-$k$ upturn of the first-order reconstruction happens because our initial density has zero shot noise but the late-time density has a small shot noise, $\bar n^{-1}=1.47\,h^{-3}\mathrm{Mpc}^3$.
This can be avoided by multiplying with $\bar t_1(k)$ given in \eqq{bart1}, which acts like a Wiener filter (thin solid line).
}
\label{fig:Pk}
\end{figure}

Having discussed the correlation of the reconstructed density with the initial conditions and the BAO signature in the power spectrum, we can also ask how well our reconstruction method recovers the full broadband power spectrum of the initial conditions.
This is shown in \fig{Pk}, where the measured power spectrum after reconstruction is divided by the linear initial power spectrum linearly scaled to the redshift $z=0$ of the simulation.

The original nonlinear power spectrum agrees with the linear power spectrum within $5\%$ at $k\le 0.11\ihMpc$ in our simulations at redshift $z=0$.
Our first-order reconstruction without transfer functions improves this slightly, so that reconstructed and linear power agree within $5\%$ at $k\le 0.16\ihMpc$.
The second-order method gives a larger improvement, agreeing with the linear power within $5\%$ at $k\le 0.28\ihMpc$.
At higher redshift, $z=0.6$, the power spectra agree with the linear one down to smaller scales, with $5\%$ agreement at $k\le 0.15\ihMpc$ for the nonlinear density, $k\le 0.33\ihMpc$ for first-order reconstruction, and $k\le 0.43\ihMpc$ for second-order reconstruction (see \fig{Pkz0pt6} in the appendix).

In our idealized simulations, our reconstruction method thus extends the $k$ range where linear theory is valid by a factor of 2 to 3. 
In principle, the reconstruction could thus enable cosmological analyses to include the power spectrum on moderately nonlinear scales, $0.15\ihMpc\lesssim k\lesssim 0.4\ihMpc$ at $z\simeq 0-0.6$, where linear theory would not be valid without reconstruction.
Since the precision of many cosmological constraints scales steeply with the number of Fourier modes included in the analysis, this could substantially improve cosmology constraints from present and future galaxy surveys.
For this to be possible in practice, however, one would have to address several complications, in particular nonlinear halo bias and redshift space distortions that affect the broadband shape of the power spectrum.

On small scales, 
$k\gtrsim 0.3\ihMpc$ at redshift $z=0$, 
the reconstruction suffers from a deficit in power relative to the linear density.
Qualitatively, this is not surprising because the reconstructed displacement field does not contain contributions from modes that underwent shell crossing, so that particles that ended up in halos are moved back farther than in the true Universe, leading to a suppression of power on small scales (see \fig{slices}).
This is the same reason for why the power spectrum in the Zeldovich approximation misses small-scale power: Particles free-stream through halos rather than undergoing nonlinear collapse to form halos, which washes out small-scale clustering power that is present in the true Universe.

More quantitatively, we show in \app{RecModel} that reconstruction fully removes nonlinear shift terms, while it only partially removes nonlinear growth and tidal terms.
While we attempt to mitigate the latter nonlinearities at second order by using the second-order reconstruction scheme, we have not attempted to remove nonshift nonlinearities at third order or higher.
The residual nonlinearities likely suppress the power spectrum on small scales.
We therefore expect the reconstructed power spectrum to agree with the linear power spectrum only on large scales, with a suppression relative to linear on smaller scales as seen in \fig{Pk}.
On smaller scales we would have to add corrections to the linear power spectrum to match the reconstructed power spectrum, in particular 1-3 and $k^2P_\mathrm{lin}$ counterterms. 
These can be computed perturbatively, and one might hope that these corrections are easier to model than the fully nonlinear power spectrum, for example because the corrections contain no BAO wiggles, but we leave this for future work.
Instead we account for the small-scale mismatch of linear and reconstructed power spectra using transfer functions $t_1(k)$ and $t_2(k)$ calibrated to the simulations as described in \app{TransferFcns}.

The fact that the reconstruction misses small-scale broadband power relative to linear is not necessarily a problem. 
All that is needed to estimate cosmological parameters is a theoretical model predicting the reconstructed power spectrum as a function of cosmological parameters. 
Such a model could be developed using e.g.~the Zeldovich approximation or higher-order LPT models.
In \app{RecModel} we provide a starting point for this, and we plan to compare this against the reconstructed density in simulations in future work (noting that 1-3 terms and a $k^2P_\mathrm{lin}$ term should also be added as mentioned above).
For standard reconstruction several models have already been developed to describe the reconstructed power spectrum \cite{Padmanabhan0812,SherwinZaldarriaga2012,Marcel1508,white1504.03677,Baldauf1504,2016MNRAS.457.2068C,Hikage1703}, and it would be interesting to extend these to our method.
To answer the question of how much cosmological information can be obtained from the full broadband shape of the reconstructed power spectrum, one additionally needs to determine the covariance between power spectrum bins after reconstruction \cite{IsoRecPan1611Fisher}.

As a slightly technical remark, we note that it is important to fill empty $\vchi$ cells with random neighbor grid cells as described in \se{RecRecipe} to avoid a large-scale bias of the power spectrum that would be caused by setting $\vchi$ to zero in empty cells (which  would effectively violate mass conservation). 
This is mainly important for the full-shape power spectrum and does not affect the correlation coefficient with initial conditions or BAO much.

\section{Conclusions}
\label{se:Conclusions}

We have presented a new method to reconstruct the initial conditions in a given volume based on recent theoretical progress relating the linear initial density with the final nonlinear large-scale structure density \cite{Baldauf:2015zga}.
We first filter the density to suppress highly nonlinear scales where perturbation theory is not valid. 
We then estimate the displacement field that relates the uniform initial density in Lagrangian space to the filtered nonlinear density in Eulerian space by iteratively moving particles back with Zeldovich displacements.
In this iterative procedure, we progressively reduce the smoothing scale to capture progressively smaller scales while keeping $\delta\lesssim 1$.
We then estimate the linear initial density as the divergence of the cumulative Zeldovich displacement.
This estimate can be improved further by adding a second-order correction.

In dark matter simulations, we find that after eight iteration steps the reconstructed density is more than $95\%$ correlated with the linear initial density of the simulation for scales $k\le 0.35\ihMpc$ at redshift $z=0$, and for $k\le 0.53\ihMpc$ at $z=0.6$.
The power in the error between initial and reconstructed fields in units of the power spectrum of the initial field, given by one minus the squared correlation coefficient between initial and reconstructed fields, is very small, 
approximately 
$1-r^2=(10^{-6}, 10^{-4}, 10^{-2})$ at $k=(0.02,0.06,0.2)\ihMpc$ at redshift $z=0$.
Our method therefore serves as an excellent estimator for the linear density.

As a concrete application, we have demonstrated that the method improves the standard reconstruction method \cite{EisensteinRec} to restore the initial linear-theory BAO signature from the final nonlinear density contrast of the Universe.
Compared to performing no reconstruction, our method improves the BAO signal-to-noise ratio by a factor of 2.7 at redshift $z=0$ and by a factor of 2.5 at $z=0.6$, matching the optimal BAO signal-to-noise ratio of the linear density in our simulated volume for both redshifts.
This improves over standard reconstruction \cite{EisensteinRec} by $70\%$ at $z=0$ and $30\%$ at $z=0.6$,
using the same idealized simulations.

This demonstrates that our method reduces large-scale flows 
to a level where their effect on the width of the BAO peak is much smaller than the intrinsic width of the linear BAO peak imprinted at recombination \cite{EisensteinSeoWhite0604}, i.e.~the iterative reconstruction works much better than it needs to for BAO measurements.
We confirmed this by showing that the reconstructed BAO scale agrees with that of the linear initial conditions in each of our simulated realizations even if cosmic variance is canceled, at least under the idealized assumptions of our simulations.
Given this overly good performance in idealized situations, the iterative reconstruction might still recover the optimal linear BAO signal-to-noise ratio in more realistic situations.
This should make our method interesting for future galaxy surveys like DESI \cite{DESIwhitepaper}, Euclid \cite{EuclidWhitePaper}, and LSST \cite{LSSTSciBook0912} that measure the BAO scale to map the expansion history of the Universe.
Similarly to other reconstruction methods, the improvement is larger toward lower redshifts where nonlinearities degrade the BAO signature more.

As a more general application, we have shown that the reconstruction also 
improves the agreement of the full broadband shape of the power spectrum with the linear power spectrum on moderately nonlinear scales.
The reconstructed and linear power spectra agree within $5\%$ for $k\lesssim 0.16\ihMpc$ at $z=0$ and $k\lesssim 0.3\ihMpc$ at $z=0.6$
if we use our first-order reconstruction method without any transfer functions.
This improves to $k\simeq 0.3\ihMpc$ at $z=0$ and $k\simeq 0.4\ihMpc$ at $z=0.6$ if we use second-order reconstruction that involves transfer functions calibrated to simulations. 
This improves the $k$ range where linear theory is valid by a factor of 2-3 relative to performing no reconstruction.
If a similar improvement can be obtained for real data, this could have substantial impact on cosmological constraints from  galaxy power spectra because the statistical uncertainty of the power spectrum drops rapidly with increasing wavenumber. 
Additional gains may be possible by modeling the shape of the reconstructed power spectrum perturbatively or by improving the reconstruction method further, but we leave this for future work.

We demonstrated that modeling the relation between the nonlinear displacement and linear density to second order and inverting it improves the correlation of the reconstruction and the initial conditions substantially on large scales. 
For the BAO application however, most of the improvement compared to previous methods is due to the improved iterative displacement field constructed in the first stage of the method.
Indeed, for BAO measurements one could also use a simple extension of the standard reconstruction method by displacing clustered and random catalogs by the improved iterative displacement as discussed in \app{ExtendedStdRec}.

In this paper, we have only studied the idealized toy model of dark matter in real space, ignoring important practical issues related to halo biasing, shot noise, redshift space distortions, and survey selection function.
These effects are expected to degrade the effectiveness of any reconstruction method because they add stochasticity and are difficult to model.  
We plan to study the quantitative impact of these effects on the reconstruction performance in future work.

Related to this, it is not entirely clear how to adopt our method to deal with real data that can have gaps and other complications.
We note, however, that the method involves only simple operations such as estimating Zeldovich displacements from the density and moving objects, and these are the same operations as for standard reconstruction \cite{EisensteinRec}.  
To apply our method to real data one could therefore try to follow the same approaches that were used to apply standard reconstruction to real data \cite{Padmanabhan2012BAORec,2014MNRAS.441.3524K}.
Using our default of eight iteration steps, our method is only a few times slower than standard reconstruction, which should be acceptable given the potential gains.

A method similar to ours has recently been proposed by Zhu \emph{et al.} \cite{IsoRecZhu1609OneD,IsoRecZhu1611ThreeD} and was studied further in \cite{IsoRecPan1611Fisher,IsoRecWang1703BAO,IsoRecYu1703Halos}, finding similar results to ours where comparisons are possible.
Their method differs from ours in mainly two aspects: (1) The displacement between initial and final conditions is estimated using a multigrid algorithm to smoothly deform coordinates (or solve the corresponding differential equation in small time steps), whereas we iteratively displace particles by their Zeldovich displacement.
(2) To estimate the linear density from the displacement, both methods first compute the divergence of the displacement, but we also add a second-order correction that improves the correlation with initial conditions.
The similar performance of the methods is likely due to the fact that the displacement field obtained by both methods improves the one-step Zeldovich displacement used by standard reconstruction, and in the regime where shell crossing is relevant, the details of the method to find the displacement likely do not matter much because the solution is not unique in that regime.
Both reconstruction methods do not assume a cosmological model.
They shall both be useful in the future because they likely enable similar gains but with different systematics because of the different operations involved.

We have demonstrated that a simple iterative reconstruction method improves significantly over the standard method in idealized simulations.  
If similar improvements can be achieved in more realistic scenarios including noise, halo biasing and redshift space distortions, this could improve BAO measurements and increase the number of power spectrum Fourier modes amenable for cosmological analysis for future galaxy surveys like DESI \cite{DESIwhitepaper}, Euclid \cite{EuclidWhitePaper}, and LSST \cite{LSSTSciBook0912}.
This could significantly improve the scientific return of these surveys, including high-precision probes of the expansion history of the Universe, dark energy, neutrino mass, and the statistics of the primordial fluctuations.

\section*{ACKNOWLEDGEMENTS}
We thank Zhejie Ding for sharing a suite of large-volume simulations optimized for BAO studies \cite{DingEtAl}.
We also thank Yu Feng for making his code base publicly available and for providing extensive help for how to use it.
We are also grateful to Zvonimir Vlah for sharing linear power spectra with and without BAO wiggles (see appendix of Ref.~\cite{Vlah1509}).
We acknowledge useful comments from An\v{z}e Slosar, Nikhil Padmanabhan, Uro\v{s} Seljak, and Marko Simonovi\'{c}.
Parts of our reconstruction algorithm use \textsf{nbodykit}, a publicly available\footnote{\url{https://github.com/bccp/nbodykit}} parallel Python postprocessing package developed by Yu Feng and Nick Hand.
Our simulations were run with the publicly available\footnote{\url{https://github.com/rainwoodman/fastpm}} \textsf{FastPM} code developed by Yu Feng \cite{fastpm}.
This research used resources of the National Energy Research Scientific Computing Center (NERSC), a DOE Office of Science User Facility supported by the Office of Science of the U.S.~Department of Energy under Contract No.~DE-AC02-05CH11231.
We also acknowledge use of the Hyperion cluster at the Institute for Advanced Study.
M.S.~acknowledges support from the Bezos Fund through the Institute for Advanced Study.

\appendix

\section{Transfer functions}
\label{app:TransferFcns}

To obtain the transfer functions needed for the second-order reconstruction, we write the reconstructed density as
\beq
\label{eq:delta0O2}
\hat \delta_0 = t_1(k) \delta_\chi(\vk) + t_2(k) \delta_\chi^{[2]}(\vk),
\eeq
where $\delta_\chi=\vnabla\cdot\vchi$ and 
\beq
\label{eq:deltachi2}
\delta_\chi^{[2]}(\vk) = \int_{\vp_1}\kappa_2(\vp_1,\vp_2) \bar t_1(p_1)\delta_\chi(\vp_1) \bar t_1(p_2)\delta_\chi(\vp_2),
\eeq
where $\vp_2=\vk-\vp_1$, $\bar t_1$ is a filter to be specified later, and $\kappa_2$ is defined in \eqq{kappa2} below.
We choose the transfer functions $t_1$ and $t_2$ in \eqq{delta0O2} to minimize the difference $\langle(\hat\delta_0-\delta_0)^2\rangle$ with the linear density in simulations.
This gives
\begin{align}
  \label{eq:16}
  t_1(k) & = \frac{1}{1-r_{12}^2}
\left(
\frac{\langle\delta_0\delta_\chi\rangle}{\langle\delta_\chi\delta_\chi\rangle}
-\frac{\langle\delta_0\delta_\chi^{[2]}\rangle}{\langle\delta_\chi^{[2]}\delta_\chi^{[2]}\rangle}
\frac{\langle\delta_\chi\delta_\chi^{[2]}\rangle}{\langle\delta_\chi\delta_\chi\rangle}
\right)
\end{align}
and
\begin{align}
  \label{eq:10}
  t_2(k) & = \frac{1}{1-r_{12}^2}
\left(
\frac{\langle\delta_0\delta_\chi^{[2]}\rangle}{\langle\delta_\chi^{[2]}\delta_\chi^{[2]}\rangle}
-\frac{\langle\delta_0\delta_\chi\rangle}{\langle\delta_\chi\delta_\chi\rangle}
\frac{\langle\delta_\chi^{[2]}\delta_\chi\rangle}{\langle\delta_\chi^{[2]}\delta_\chi^{[2]}\rangle}
\right),
\end{align}
where we assumed $\bar t_1$ to be fixed, and we used the correlation coefficient of the first and second order contribution, 
\begin{align}
  \label{eq:9}
r_{12}^2(k) = \frac{\langle\delta_\chi\delta_\chi^{[2]}\rangle^2}{\langle\delta_\chi\delta_\chi\rangle\langle\delta_\chi^{[2]}\delta_\chi^{[2]}\rangle}.
\end{align}

To specify the transfer function $\bar t_1$ inside the second order contribution in \eqq{deltachi2} we minimize $\langle(\bar t_1\delta_\chi-\delta_0)^2\rangle$, i.e.~the error in the absence of a second order contribution.
This gives
\begin{align}
  \label{eq:bart1}
 \bar t_1(k)=\frac{\langle\delta_0\delta_\chi\rangle}{\langle\delta_\chi\delta_\chi\rangle},
\end{align}
Indeed, we have $\bar t_1=t_1$ if $\langle\delta_\chi\delta_\chi^{[2]}\rangle$ were to vanish.
On large scales $\delta_\chi$ agrees with $\delta_0$ so that $\bar t_1=1$.
On smaller scales $\delta_\chi$ differs from $\delta_0$ so that $\bar t_1$ approaches zero, which down-weights small-scale modes in \eqq{deltachi2}.
Using $\bar t_1$, we can rewrite $t_2$ in a form reminiscent of Gram-Schmidt orthogonalization,
\begin{align}
  \label{eq:11}
  t_2(k) = \frac{1}{1-r_{12}^2} \frac{\la(\delta_0-\bar t_1\delta_\chi),\delta_\chi^{[2]}\ra}{\langle\delta_\chi^{[2]}\delta_\chi^{[2]}\rangle}.
\end{align}

In the above equations, the shorthand notation $\la\delta_0\delta_\chi\ra\equiv P_{\delta_0,\delta_\chi}(k)$ refers to the cross spectrum between $\delta_0$ and $\delta_\chi$ as a function of wavenumber $k$, and analogously for other fields.
All transfer functions depend only on the modulus $k$ of wave vectors and can therefore be calibrated from simulations.

\begin{figure}[tbhp]
\includegraphics[width=0.45\textwidth]{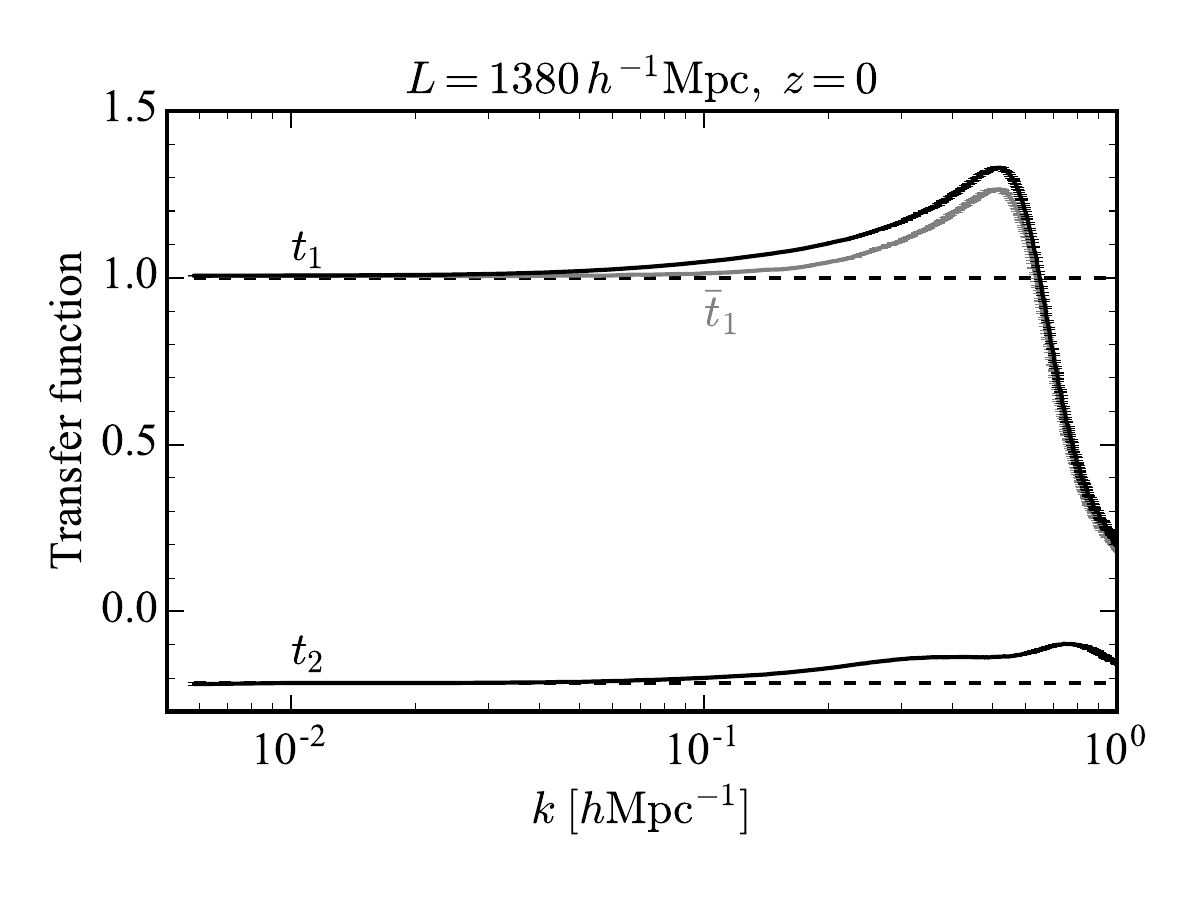}
\includegraphics[width=0.45\textwidth]{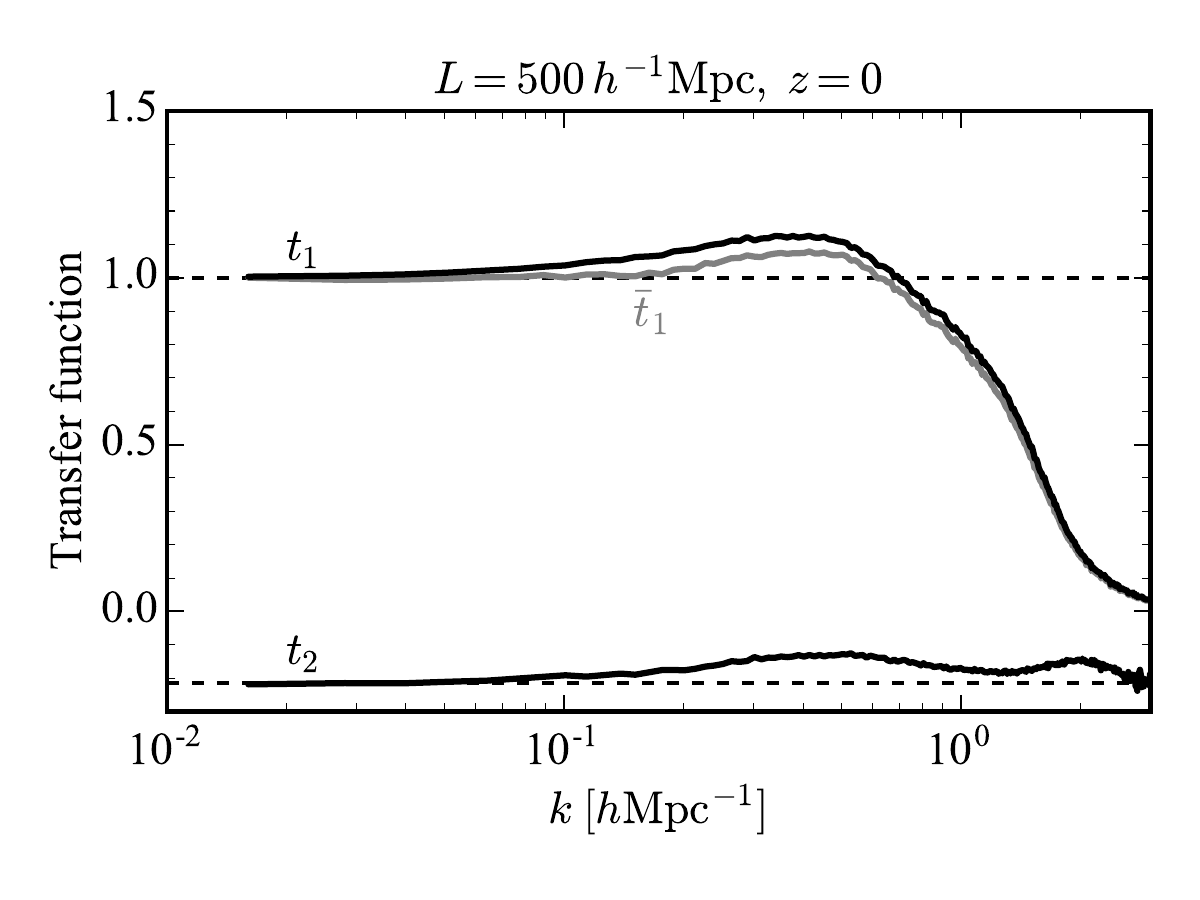}
\caption{\emph{Upper panel:} Transfer functions measured from power spectra of ten large-volume simulations with $L=1380\hMpc$, with displacement field from eight iteration steps and minimum smoothing scale $R_\mathrm{min}=1.01\,L/N_\mathrm{grid}=2.7\hMpc$.
\emph{Lower panel:} Same but for a single smaller simulation with $L=500\hMpc$, and $R_\mathrm{min}=1.01\,L/N_\mathrm{grid}=0.98\hMpc$. 
The transfer functions are computed from cross spectra with the initial conditions of the simulations as described in \app{TransferFcns}.
The spectra are raw spectra measured on $512^3$ grids without correcting for shot noise or CIC effects that are relevant at high $k$.
Dashed lines show theoretical low-$k$ limits.
The transfer functions could be modeled analytically but we did not attempt this here.
 }
\label{fig:TransferFcns}
\end{figure}

To calibrate the transfer functions, we measure the auto- and cross-spectra of $\delta_0$, $\delta_\chi$ and $\delta_\chi^{[2]}$ from the simulations, compute the transfer functions, and average over realizations where more than one realization is available. The result is shown in \fig{TransferFcns}.

At low $k$, the transfer functions approach their theoretical low-$k$ limits given by
\begin{align}
  \label{eq:1}
  \lim_{k\rightarrow 0}t_1(k) = \lim_{k\rightarrow 0}\bar t_1(k) = 1
\end{align}
and
\begin{align}
  \label{eq:2}
  \lim_{k\rightarrow 0}t_2(k) = -\frac{3}{14}.
\end{align}
These limits follow by using the expansion \eq{ExpandDeltaChiInDelta0} of $\delta_\chi$ in $\delta_0$, modeling all spectra at leading order in $P_\mathrm{lin}$, and using the large-scale limit given in \eqq{Fchi2limit}.
At higher $k$, the transfer functions deviate from that limit, but they are still rather smooth functions of $k$.
It should therefore be possible to either model them analytically or continue the practice of calibrating them against simulations.
Ultimately, the transfer functions should also include halo bias and redshift space distortions.

\section{Extended standard reconstruction}
\label{app:ExtendedStdRec}

In this appendix we discuss a simple extension of the standard reconstruction method of \cite{EisensteinRec} that could be used as an alternative to the reconstruction method discussed in the main text.
We regard the method in the main text as superior on theoretical grounds because the conversion from nonlinear displacement to linear density is better motivated, but the alternative method discussed here could in principle be useful for practical purposes (for example, it might be easier to deal with the survey selection function).

\subsection{Method}

We first apply the iterative procedure described in \se{RecRecipe} to find the displacement field $\vchi^{(8)}$ with eight iteration steps, which improves the Zeldovich displacement used in standard reconstruction.
This is the same as the first stage of the method described in the main text.
We can now ask how well we can do if we apply the second stage of standard reconstruction to the improved displacement field.
Following \se{StdRecRecipe}, we thus displace the clustered catalog and a uniform catalog by $\vchi^{(8)}$ defined in Eulerian coordinates $\vx$ and then take the density difference.
This second stage differs from the method in the main text, where we take the divergence of $\vchi^{(8)}$ defined in approximate Lagrangian coordinates $\hat\vq$ with an optional second order correction; also see Table~\ref{tab:CompareAlgorithms}.

This extension of the standard method based on an iterative displacement field is very similar to the iterative standard reconstruction of Ref.~\cite{SeoBAOShift}, where no improvement over standard reconstruction was found.
However, an important difference is that the smoothing scale used to compute the iterative Zeldovich displacements was held fixed in Ref.~\cite{SeoBAOShift}, whereas we progressively decrease the smoothing scale from step to step to access progressively smaller scales.  
We believe that this is the main difference compared to the implementation in \cite{SeoBAOShift}, but there may be additional differences.  
As shown in the lowest row in Table~\ref{tab:RecParamsComparison}, we also find little improvement from iterations if we keep the smoothing scale fixed, consistent with the findings of \cite{SeoBAOShift}, but, as we discuss next, we find substantial improvements when reducing the smoothing scale from step to step, $\epsilon_R<1$.

\subsection{Performance and discussion}

\begin{figure*}[tbhp]
\includegraphics[width=0.9\textwidth]{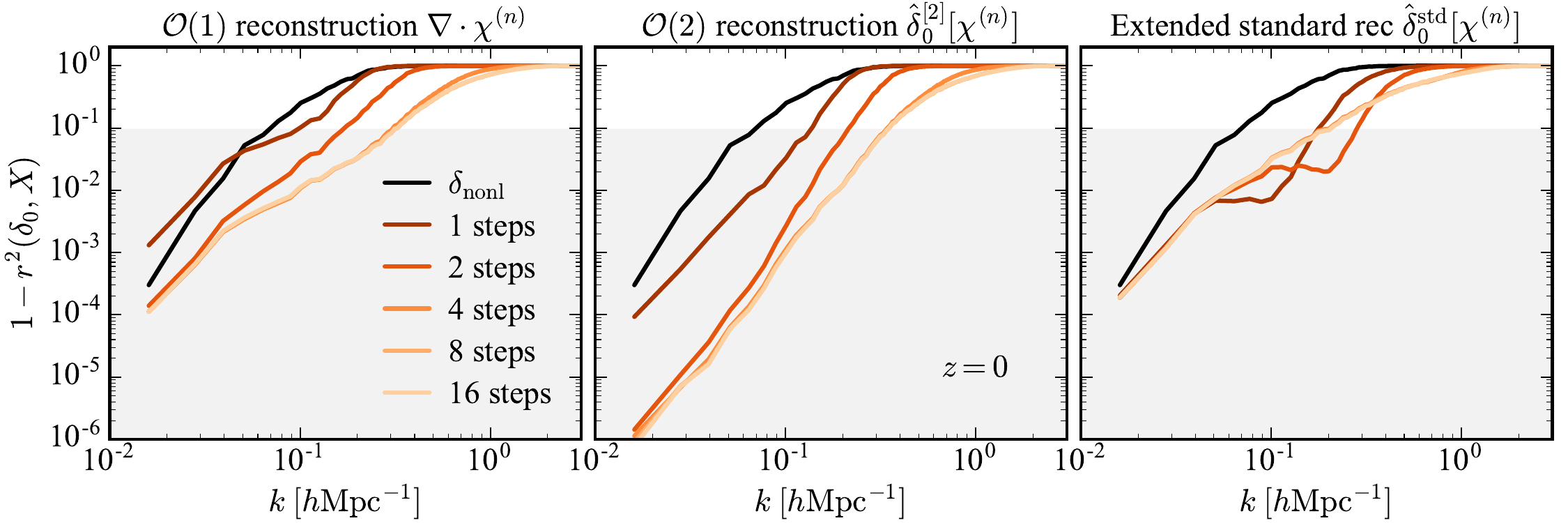}
\caption{Power of the error of the reconstructed density relative to the linear density, represented by one minus the squared correlation coefficient with the linear initial conditions. 
In the grey shaded area, the reconstructed density is more than $95\%$ correlated with the linear density.
Brighter colors represent a better displacement field $\vchi$ obtained by running more iteration steps (1, 2, 4, 8, or 16).
The curves are from our $L=500\hMpc$ simulation at $z=0$.
}
\label{fig:1mr2Steps}
\end{figure*}

The right panel of \fig{1mr2Steps} shows that after applying the extended standard reconstruction described above, the large-scale density is very correlated with the linear initial conditions.
The achieved correlation is similar to the $\mathcal{O}(1)$ reconstruction discussed in the main text and shown in the left panel of \fig{1mr2Steps}, but worse than the $\mathcal{O}(2)$ reconstruction shown in the middle panel of \fig{1mr2Steps}.

This demonstrates that the second order scheme of \eqq{hat_delta0_from_2ndorder} is more successful at converting a given nonlinear displacement field to the linear density than just taking the divergence of that displacement or displacing clustered and random catalogs and taking their density difference.
This is expected given that the second order scheme inverts the model of \eqq{deltaChiAsFcnOfDelta0} which has been validated against simulations \cite{Baldauf:2015zga}.

For measuring the BAO scale, reconstruction does not need to work perfectly as discussed in \cite{EisensteinRec} and \se{ResidualShifts}.
Indeed, Tables~\ref{tab:BAOFitsMeanAppdx} and \ref{tab:BAOFitsScatterAppdx} show that the extended standard reconstruction works sufficiently well that it fully recovers the linear BAO signal-to-noise ratio, improving significantly over the standard one-step reconstruction.
For BAO measurements, the $\mathcal{O}(1)$ and $\mathcal{O}(2)$ methods from the main text and the extended standard method discussed in this appendix thus all perform equally well, improving the standard method based on a one-step displacement.

This result shows that the main improvement compared to the standard one-step method of \cite{EisensteinRec} comes from the improved iterative displacement field.
To obtain this, it is crucial to progressively reduce the smoothing scale in the iterations to access smaller and smaller scales with each iteration step (see Table~\ref{tab:RecParamsComparison}).
Once this improved displacement field is obtained in the first stage of reconstruction, it does not matter for BAO measurements how the displacement is converted to an estimate for the linear density in the second stage.
For BAO measurements from real data one should therefore choose the second stage based on practical considerations and characteristics of the given data set.
For applications beyond BAO measurements, the second order reconstruction in the main text should be used instead because it achieves the highest correlation with the initial conditions and best broadband power spectrum shape.

\begin{table}[tbhp]
\centering
\renewcommand{\arraystretch}{1.0}
\begin{tabular}{@{}p{2.2cm}lllll@{}}
\toprule
& \phantom{} & \multicolumn{3}{l}{Mean BAO scale}  \\
Field && vs lin.~theory &\phantom{$\,$}& vs lin.~realization \\
\colrule
Ext std rec && $+0.02\,\mathrm{Mpc}$ [$+0.01\%$] && $-0.03\,\mathrm{Mpc}$ [$-0.02\%$] \\
Scaled $\mathcal{O}(1)$ rec && $+0.06\,\mathrm{Mpc}$ [$+0.04\%$] && $+0.01\,\mathrm{Mpc}$ [$+0.01\%$] \\
\botrule
\end{tabular}
\caption{Systematic bias of the best-fit BAO scale at $z=0$ as in Table~\ref{tab:BAOFitsMean} but for the extended standard reconstruction ('Ext std rec') method of \app{ExtendedStdRec} and for the scaled $\mathcal{O}(1)$ reconstruction given by $t_1(k)\delta_\chi(\vk)$, where $\delta_\chi=\vnabla\cdot\vchi^{(8)}$, and $t_1$ is from \app{TransferFcns}. There is no evidence for a systematic BAO bias for any of the methods.
}
\label{tab:BAOFitsMeanAppdx}
\end{table}

\begin{table}[thp]
\centering
\renewcommand{\arraystretch}{1.0}
\begin{tabular}{@{}p{2.2cm}lllll@{}}
\toprule
& \phantom{} & \multicolumn{3}{l}{Rms scatter of BAO scale} \\
Field && vs lin.~theory &\phantom{$\quad$}& vs lin.~realization \\
\colrule
Ext std rec && $0.31\,\mathrm{Mpc}$ [$0.21\%$] && $0.18\,\mathrm{Mpc}$ [$0.12\%$] \\
Scaled $\mathcal{O}(1)$ rec && $0.4\,\mathrm{Mpc}$ [$0.27\%$] && $0.11\,\mathrm{Mpc}$ [$0.07\%$] \\
\botrule
\end{tabular}
\caption{Rms scatter of the best-fit BAO scale at $z=0$ as in Table~\ref{tab:BAOFitsScatter} but for the extended standard reconstruction method of \app{ExtendedStdRec} and for the scaled $\mathcal{O}(1)$ reconstruction given by $t_1(k)\delta_\chi(\vk)$. The left column shows that in both cases the BAO uncertainty is consistent with that of the BAO scale in the initial conditions ($0.24\%$) within the uncertainty of the ten simulations. The right column represents the nonlinear noise contribution due to residual shift terms after reconstruction as discussed in \se{ResidualShifts}.
For BAO, the methods are thus comparable to the method described in the main text.
}
\label{tab:BAOFitsScatterAppdx}
\end{table}

\section{Fitting BAO from the power spectrum}
\label{app:BAOFitting}
We employ the following procedure to estimate the BAO scale from the power spectra measured in our simulations before and after reconstruction.
In each realization, we compute the fractional BAO signal of the power spectrum, $\hat d(k)=\hat P_\mathrm{wiggle}(k)/\langle\hat P_\mathrm{nowiggle}(k)\rangle-1$. 
Here, $\hat P_\mathrm{wiggle}$ is the power spectrum measured in a simulation realization with BAO wiggles; $\hat P_\mathrm{nowiggle}$ is the same but in a simulation initialized with no BAO wiggles; and $\langle\cdot\rangle$ represents the average over our ten realizations.
We then fit that data in each realization with a simple theoretical model that multiplies the BAO scale by a factor $\alpha$ and suppresses the BAO wiggles using a damping factor $\Sigma$,
\begin{align}
  \label{eq:1}
m(k) = e^{-(k\Sigma)^2/2}\left[O(k/\alpha)-1\right].
\end{align}
Here $O(k)=P^\mathrm{lin}_\mathrm{wiggle}(k)/P^\mathrm{lin}_\mathrm{nowiggle}(k)$ is the ratio of the theoretical linear wiggle and nowiggle power spectra that were used to initialize the simulations.
We fit for $\alpha$ and $\Sigma$ by minimizing the chi-squared $\sum_k[\hat d(k)-m(k)]^2/\sigma^2(k)$ at $0<k\le k_\mathrm{max}^\mathrm{fit}=0.6\ihMpc$ for every realization.\footnote{We allow $\Sigma$ to vary between $0$ and $20\hMpc$, and $\alpha$ between $0.9$ to $1.1$. This is conservative given that $\alpha$ should deviate by less than $1\%$ from unity in any realization of the large-volume simulations.
Best-fit values are consistent when fitting with scipy.optimize.minimize's Nelder-Mead, Powell or TNC algorithms.
} 
This gives the best-fit BAO scale in each realization, $\hat r_\mathrm{BAO}=\hat\alpha\, r_\mathrm{BAO}^\mathrm{fid}$, as shown in \fig{histalpha}.
We estimate the uncertainty of the measured BAO scale by computing the scatter of the best-fit BAO scale between the realizations.
This Monte Carlo method to estimate the BAO uncertainty should provide a robust estimate of the true uncertainty because it quantifies how much the BAO scale estimated from our particular fitting procedure varies among different realizations of the Universe.

When fitting for the BAO scale, we assume a Gaussian diagonal covariance $\mathrm{cov}(\hat d(k),\hat d(k')) = \delta_{kk'} \sigma^2(k)$ with
\begin{align}
  \label{eq:3}
\sigma^2(k) = \frac{2}{N_\mathrm{modes}(k)}
\la\frac{[\hat P_\mathrm{wiggle}(k)]^2}{\langle\hat P_\mathrm{nowiggle}(k)\rangle^2}\ra
\end{align}
estimated from the simulations.
This should be regarded just as a particular weight in the fitting process to up-weight the power spectrum on small scales where there are more modes.
If the assumed Gaussian covariance is wrong, the estimator for the BAO scale is suboptimal because it employs a suboptimal weight.  In that case, the uncertainty estimated from the scatter of the best-fit BAO scale between realizations overestimates the uncertainty compared to a more optimal estimator or fitting procedure based on the correct covariance.
As a consequence, our estimated BAO uncertainty is a conservative estimate of the true uncertainty.

\section{Results at redshift \texorpdfstring{$z=0.6$}{z=0.6}}
\label{app:HighZ}

\begin{figure}[tp]
\includegraphics[width=0.48\textwidth]{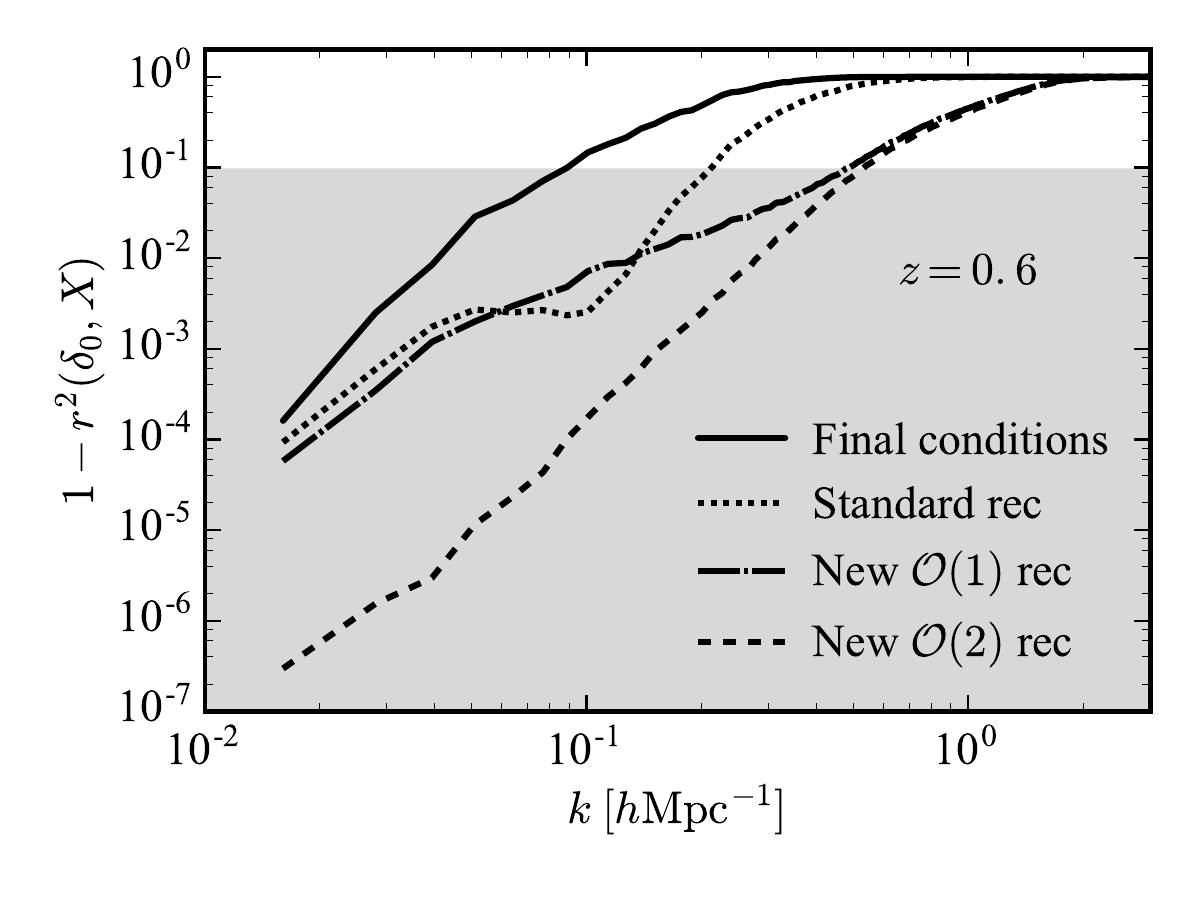}
\caption{Same as \fig{rccSimple} but at redshift $z=0.6$.
The new reconstruction is more than $95\%$ correlated with the initial conditions at $k\le 0.48\ihMpc$, or at $k\le 0.53\ihMpc$ if second order corrections are included in the method.
For comparison, the wavenumber where the correlation with initial conditions drops below $95\%$ is $k=0.21\ihMpc$ for standard reconstruction, and $k=0.09\ihMpc$ for the nonlinear density without reconstruction in our setup.
}
\label{fig:rccSimplez0pt6}
\end{figure}

\begin{figure}[tp]
\includegraphics[width=0.48\textwidth]{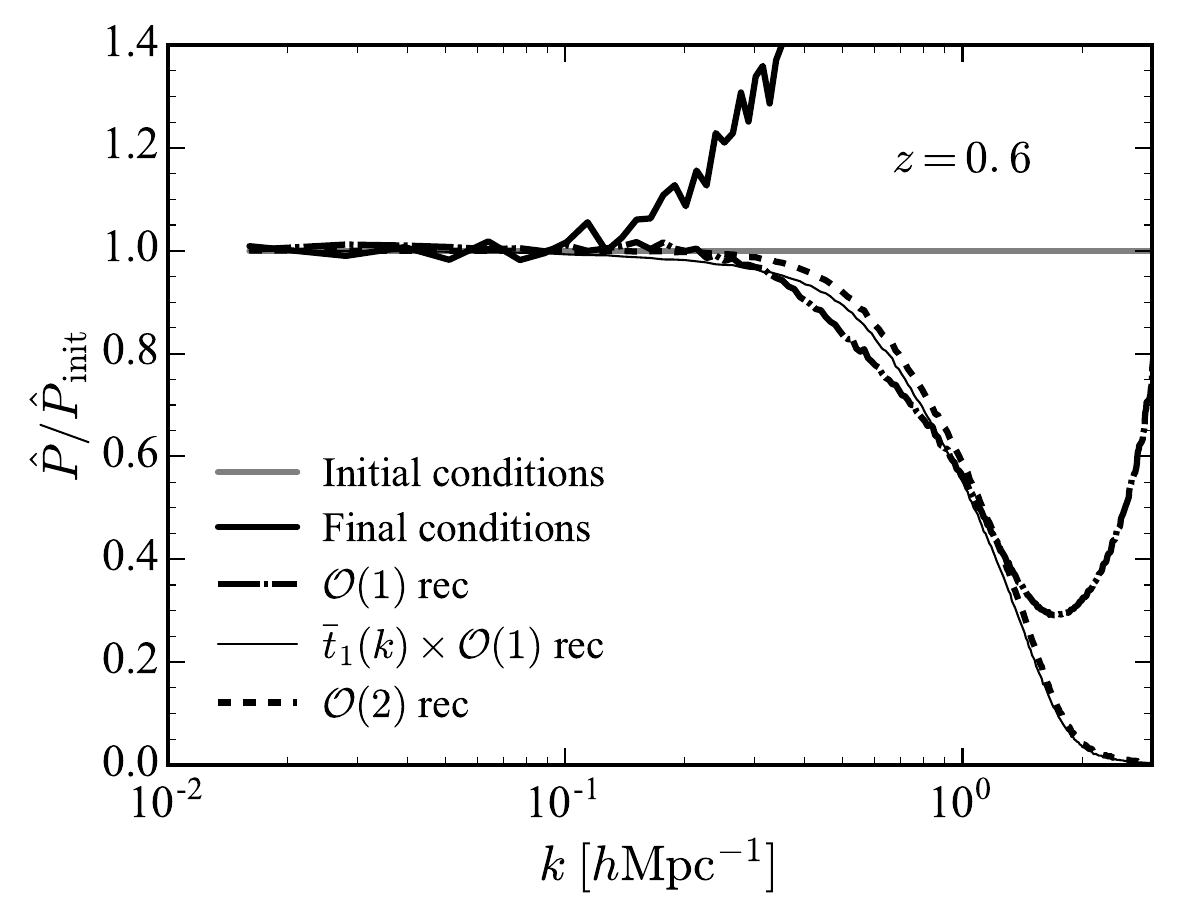}
\caption{Same as \fig{Pk} but at redshift $z=0.6$.
}
\label{fig:Pkz0pt6}
\end{figure}

Figures~\ref{fig:rccSimplez0pt6} and \ref{fig:Pkz0pt6} show reconstruction results at redshift $z=0.6$ as opposed to the redshift $z=0$ that was used in the main text.  
The densities before and after reconstruction match the linear density better at this higher redshift than at lower redshift, which is as expected because nonlinearities are smaller at higher redshift for any given scale.

\section{Parameters and convergence tests}
\label{app:TechnicalDetails}

In this appendix we discuss some choices we made for the reconstruction parameters, and some basic convergence tests of our simulations.

\subsection{Reconstruction parameters}
\label{app:recparamsAppendix}

\begin{table*}[tp]
\centering
\renewcommand{\arraystretch}{1.0}
\begin{tabular}{@{}rrrrrrrrrrrrrrr@{}}
\toprule
&&& \phantom{$\quad$} & \multicolumn{3}{l}{2 steps} & \phantom{$\qquad$} & \multicolumn{3}{l}{8 steps} & \phantom{$\qquad$} & \multicolumn{3}{l}{16 steps} \\
\multicolumn{1}{r}{$R$} & \multicolumn{1}{r}{$\epsilon_R$} & \multicolumn{1}{r}{$\epsilon_s$}
&& \multicolumn{1}{l}{$k=0.1$} && \multicolumn{1}{l}{$k=0.6$}
&& \multicolumn{1}{l}{$k=0.1$} && \multicolumn{1}{l}{$k=0.6$}
&& \multicolumn{1}{l}{$k=0.1$} && \multicolumn{1}{l}{$k=0.6$} \\
\colrule
5 & 0.5 & \multicolumn{1}{r|}{0.5} && $5.9\times 10^{-3}$ && 0.97 && $4.4\times 10^{-4}$ && 0.19 && $3.0\times 10^{-4}$ && 0.17 \\
10 & 0.5 & \multicolumn{1}{r|}{0.5} && $6.5\times 10^{-3}$ && 0.9 && $4.1\times 10^{-4}$ && 0.19 && $2.7\times 10^{-4}$ && 0.17 \\
10 & 0.5 & \multicolumn{1}{r|}{1} && $6.1\times 10^{-3}$ && 0.94 && $4.0\times 10^{-4}$ && 0.19 && $2.9\times 10^{-4}$ && 0.18 \\
20 & 0.5 & \multicolumn{1}{r|}{1} && $1.9\times 10^{-2}$ && 0.94 && $4.1\times 10^{-4}$ && 0.2 && $2.9\times 10^{-4}$ && 0.18 \\
1 & 1 & \multicolumn{1}{r|}{0.33} && $2.2\times 10^{-2}$ && 0.86 && $3.0\times 10^{-3}$ && 0.33 && $3.4\times 10^{-3}$ && 0.31 \\
10 & 1 & \multicolumn{1}{r|}{0.5} && $8.4\times 10^{-3}$ && 0.92 && $4.2\times 10^{-3}$ && 0.92 && $4.0\times 10^{-3}$ && 0.92 \\
10 & 1 & \multicolumn{1}{r|}{1} && $8.6\times 10^{-3}$ && 0.93 && $4.2\times 10^{-3}$ && 0.93 && $4.1\times 10^{-3}$ && 0.93 \\
\botrule
\end{tabular}
\caption{Performance of the second-order reconstruction for some choices of the initial smoothing scale $R$, the reduction factor $\epsilon_R$, and the displacement amplitude $\epsilon_s$ (all other parameters are set to their default values described in \se{recparams}).
The table shows one minus the squared correlation coefficient between reconstructed and linear density, $1-r^2(\delta_\mathrm{0},\delta_\mathrm{rec})$, after 2, 8 and 16 iteration steps, at $k=0.1\ihMpc$ and $k=0.6\ihMpc$, in our $L=500\ihMpc$ simulation at $z=0.6$.
Lower numbers correspond to better reconstruction. 
Going from 2 to 8 steps improves performance substantially, while more than 8 steps yields diminishing returns.
After 8 or 16 steps steps, all reconstruction algorithms with smoothing reduction factor $\epsilon_R=0.5$ perform similarly well.
(Note that the results of the table were obtained using suboptimal transfer functions, which is why results for $(R,\epsilon_R,\epsilon_s)=(10,0.5,1)$ differ slightly from \fig{rccSimplez0pt6}; the overall trends should not depend on this though.)
}
\label{tab:RecParamsComparison}
\end{table*}

Our reconstruction algorithm has several parameters as described in \se{recparams}.
Table~\ref{tab:RecParamsComparison} shows results for different parameter choices.
This demonstrates that the final performance of the method is relatively insensitive to the detailed parameter values.
Some qualitative choices are important though:
It is important to start with a relatively large smoothing scale, $R_\mathrm{init}\gtrsim 5\hMpc$, so that the smoothed overdensity is less than unity for most grid points and the Zeldovich approximation is applicable.
The smoothing scale should also decrease from one iteration to the next, $\epsilon_R<1$, to reconstruct progressively smaller scales in the iteration.
However, since reconstruction likely becomes inefficient on very small scales where shell crossing dominates, we stop decreasing the smoothing scale at $R_\mathrm{min}$. A reasonable choice may be $R_\mathrm{min}\sim 1\hMpc$.
We work with $R_\mathrm{min}=1.01\,L/N_\mathrm{grid}$ throughout, which gives $R_\mathrm{min}=0.99\hMpc$ for our small-volume simulation and $R_\mathrm{min}=2.7\hMpc$ for our large-volume simulations.

For the other parameters, we can use simple heuristics. 
The number of iteration steps for the displacement $\vchi$ can be determined by monitoring the final quantity of interest and stopping the iteration once that quantity stops changing significantly.
\fig{1mr2Steps} demonstrates that less than ten iteration steps should be sufficient to achieve convergence for most applications.
The size of the regular grid should be chosen such that the smallest length scale of interest is still resolved by the grid. 
We work with $N_\mathrm{grid}^3=512^3$ grid points throughout the postprocessing of the simulations, corresponding to a grid resolution of $\Delta x=0.98\hMpc$ for the $L=500\hMpc$ simulations, and $\Delta x=2.7\hMpc$ for the $L=1380\hMpc$ simulations that we use for studying the BAO scale.
To reduce potential aliasing effects, we truncate small-scale modes with $k> k_\mathrm{max}=2\pi/L\times N_\mathrm{grid}/2$.

The optional displacement factor $\epsilon_s$ can be set to less than unity to avoid potential overshooting when displacing objects, for example if the overdensity is large and the Zeldovich displacement may not be appropriate. 
However, if the initial smoothing scale is chosen sufficiently large, for example, $R_\mathrm{init}\gtrsim 5\hMpc$ at $z=0.6$, the overdensity tends to be less than unity and full displacements with $\epsilon_s=1$ seem to work well.

\subsection{Convergence of simulations}
As a basic check for convergence of the \textsf{FastPM} simulations, we ran a simulation with $40$ time steps linearly spaced between $a=0.1$ and $a=1$, and a second more accurate simulation with $120$ time steps linearly spaced between $a=0.01$ and $a=1$. 
Both simulations used $2048^3$ particles, box size $L=500\hMpc$, and we apply reconstruction to a $1\%$ dark matter subsample at $z=0.6$.
The correlation coefficient of the reconstructed density with the initial conditions differs by less than $1\%$ between these two simulations, for all reconstruction methods considered in this paper, and for any number of iterations steps used in the reconstruction procedure.
This indicates that the simulations have converged in the sense that the final result is robust against changes of starting time and number of time steps used to run the simulations.

\section{Modeling the new reconstruction method}
\label{app:RecModel}

Given a prescription for reconstruction, we can try to model the statistics of the reconstructed density, similar to previous efforts \cite{Padmanabhan0812,SherwinZaldarriaga2012,Marcel1508,white1504.03677,Baldauf1504,2016MNRAS.457.2068C,Hikage1703} modeling the density after the standard reconstruction of \cite{EisensteinRec}.

\subsection{Undoing shift terms}

From the mapping between Lagrangian and Eulerian space $\vec x=\vec q+\vec \psi$ and mass conservation
\beq\label{measure}
\bar \rho \ d^3\vq = \bar\rho (1+\delta) \ d^3\vx,
\eeq
we have that $1+\delta=1/J_q$, where $J_q$ is the determinant of the Jacobian matrix
\beq
A_{ij}\equiv\frac{\partial x_i}{\partial q^j}=\delta^K_{ij}+\psi_{i,j}
\eeq
where $\delta^K_{ij}$ is the Kronecker delta (not to be confused with density contrast). Spatial indices are always raised and lowered using $\delta^K_{ij}$ and its inverse. When convenient we will use a comma to denote spatial derivatives. 
We can obtain the final density directly from $J_q$ which we have solved explicitly in terms of $\vq$. Using \eqref{measure}
\beq\label{deltaJ}
\delta (\vx) = \Big({1\over J_q(\vq)} -1\Big)_{\vx=\vq+\vps} 
\eeq
This expression should be used perturbatively expanding up to a given order in the displacement field. 

$J_q$ depends only on displacement gradients $\psi_{i,j}$ while there are nonlinear shift terms from the mapping from $\vq$ to $\vx=\vq+\vps$ that depend on the displacement $\psi_i$ itself. These two sets of terms can have different sizes.
The shift terms can be large and should be resummed at the BAO scale if one wants to get a good estimate of the correlation function \cite{SenatoreZaldarriaga1404,Baldauf1504}; even a linear long-wavelength component of the displacement can move particles over a large enough distance that truncating a perturbative expansion of the form $f(\vq+\vps)=f(\vq)+\vps\cdot\vnabla f+\cdots$ would lead to large errors. The goal of reconstruction is to undo the shift terms so that the nonlinearities arising from these shift terms are minimized.

\subsection{Density of particles shifted by \texorpdfstring{$\vchi$}{chi}}

Given a displacement field $\vc$, let us make a coordinate transformation from the final Eulerian coordinates $\vx$ to new coordinates $\hat\vq$ defined by
\beq
\vx=\hat\vq+\vc(\hat\vq).
\eeq
If the displacement $\vc$ were to coincide with the true Lagrangian-to-Eulerian displacement $\vps$, then the $\hat\vq$ coordinates would equal the true Lagrangian $\vq$ coordinates. In that case the density of particles in $\hat\vq$ space would be uniform. 
If $\vc$ is an approximate estimate of the true displacement $\vps$, then $\hat\vq$ are estimated approximate Lagrangian coordinates, and the density of particles in $\hat\vq$ space is approximately uniform (the level to which they are uniform is shown in the upper panel of \fig{PowerOfDisplacedCat} that is based on progressively better $\vchi$).

Similar to above, we can use mass conservation to relate the density of particles in $\hat\vq$ and $\vx$ spaces,
\beq\label{measurey}
(1+\delta_{\hat q}) \ d^3\hat\vq =  (1+\delta) \ d^3\vx,
\eeq
so that
\beq
(1+\delta_{\hat q}) = [1+\delta(\vx(\hat\vq))] J_{\hat q}(\hat \vq),
\eeq
where $J_{\hat q}$ is the determinant of the Jacobian matrix
\beq
B_{ij}\equiv\frac{\partial x_i}{\partial \hat q^j}=\delta_{ij}^K+\chi_{i,j}.
\eeq

We can now compute the Fourier transform of the density field in Fourier space $\Delta (\vk)$:  
\bea\label{defdelta}
\Delta (\vk)&\equiv& \int d^3 \hat q \ [1+\delta_{\hat q}(\hat\vq)] e^{i\vk\cdot \hat\vq} \nonumber \\
&=&  \int d^3 \hat q \ [1+\delta(\vx(\hat\vq))] J_{\hat q}(\hat\vq)  e^{i\vk\cdot \hat\vq} \nonumber \\
&=& \int d^3 x \ [1+\delta(\vx)]  e^{i\vk\cdot (\vx - \vc)} \nonumber \\
&=& \int d^3 q \    e^{i\vk\cdot[ \vq + \vps(\vq) - \vc]}, 
\eea
where in the last two expressions $\vc$ should be thought of as a function of $\vx$ and $\vq$, respectively,  through the relation
\beq
\hat\vq + \vc = \vx = \vq + \vps(\vq).
\eeq
Ideally the density in Lagrangian space would be uniform and thus $\Delta (\vk)$ would be zero for nonzero $\vk$. The density is slightly nonuniform if the estimated and true Lagrangian coordinates differ, $\hat\vq\neq \vq$, which is the case if the estimated and true displacements differ, $\vc\neq\vps$.

Given $\vc$ one can construct the density in $\hat\vq$ space and measure $\Delta (\vk)$ directly from the data. The goal of this appendix is to compute the statistics of  $\Delta (\vk)$. One should keep in mind that $\vc$ will also be computed from the data itself, it will be a function of $\delta$. 
In the EFT one has a perturbative expansion for $\vps$ and $\delta$.

\subsection{Displacement field}

We can obtain an  expression for $\vc$ if we demand that the density field computed using the Zeldovich approximation starting with the displacement $\vc$ equaled the filtered version of the density. The Zeldovich approximation has all the nonlinearities from the displacements but no dynamical interaction among the different modes. We are basically trying to solve for the nonlinear displacement, of course once one is inside the nonlinear regime, this is not really possible. That is to say we will find a displacement field that produces the final density from a uniform distribution, but this is not unique as we can always exchange particles after we are done in any given solution and create a new solution. Thus what we get on nonlinear scales is a bit random. 

\begin{widetext}

If we equated the Zeldovich density to the filtered one, we would obtain:
\beq
i \vk\cdot\vc(\vk) +  \sum_{n=2}^{\infty}\int_{\vp_1,\dots,\vp_{n-1}} F_Z(\vp_1,\cdots ,\vp_n)  [i \vp_1\cdot\vc(\vp_1)] \cdots [i \vp_n\cdot\vc(\vp_n)] = W(k) \delta(\vk),
\eeq
where $\vk = \vp_1 + \cdots + \vp_n$ and $F_Z$ are the kernels of the Zeldovich approximation:
\beq
F_Z(\vp_1,\cdots ,\vp_n) = {1 \over n !} {\vk \cdot \vp_1 \over p_1^2} \cdots {\vk \cdot \vp_n \over p_n^2}.
\eeq

We then solve the equation perturbatively to obtain:
\beq
\delta_\chi\equiv i \vk\cdot\vc(\vk)  = W(k) \delta(\vk) -{1\over 2}  {\vk \cdot \vp_1 \over p_1^2 } {\vk \cdot \vp_2 \over p_2^2 }  W(p_1) \delta(\vp_1) W(p_2) \delta(\vp_2) + \cdots. 
\eeq
In general we have:
\beq
\delta_\chi=\sum_{n=1}^{\infty} F_Z^{-1}(\vp_1,\cdots ,\vp_n) W(p_1) \delta(\vp_1)\cdots  W(p_n) \delta(\vp_n)
\eeq
The perturbative version of the solution has
\bea
 F_Z^{-1}(\vp_1) &=& 1 \nonumber \\
 F_Z^{-1}(\vp_1,\vp_2) &=& -F_Z(\vp_1,\vp_2) \nonumber \\
  F_Z^{-1}(\vp_1,\vp_2,\vp_3) &=& -F_Z(\vp_1,\vp_2,\vp_3) + 2 [F_Z(\vp_1,\vp_2) F_Z(\vp_1+\vp_2,\vp_3)]_{sym},
\eea
where $[...]_\mathrm{sym}$ stands for symmetrized in the momenta. 
In standard reconstruction one is just using the linear version of this equation:
\beq
\delta_\chi \approx  W  \delta.
\eeq

\subsection{Flavors of reconstruction}

In standard reconstruction a uniform field in $\vx$ space is generated which is then shifted with the same displacement $\vc$. Let us call this field $\Delta_S$, 
\bea\label{defdeltaS}
\Delta_S (\vk)&\equiv& \int d^3 \hat q \ (1+\delta_{\hat q}(\hat\vq)) e^{i\vk\cdot \hat\vq} \nonumber \\
&=&  \int d^3 \hat q \  J_{\hat q}(\hat\vq)  e^{i\vk\cdot \hat\vq} \nonumber \\
&=& \int d^3 x \   e^{i\vk\cdot (\vx - \vc)}, 
\eea
where we have used the same expressions used to compute $\Delta(\vk)$ but in this case the overdensity in $\vx$ space is zero. 
In the standard reconstruction algorithm, one estimates $\delta_0$ as:
\bea
\hat \delta_0^{S}(\vk) &=&  \Delta(\vk) - \Delta_S (\vk) \nonumber\\
&=&  \int d^3 \hat q \ \delta(\vx(\hat\vq)) J_{\hat q}(\hat\vq)  e^{i\vk\cdot \hat\vq} 
\eea

If one has iteratively or perturbatively found a displacement field that recovers the density one can perhaps use directly that field as the estimate of the linear density field. This is what we do in this paper, 
\beq
\hat \delta_0^{N}(\vk) = {\delta_\chi(\vk) }.  
\eeq
We could also divide by $W(k)$ to undo the filtering so as to recover the linear density field at lowest order. 

We compare with standard reconstruction in the main text as well as in \app{ExtendedStdRec}. Of course our new estimate of the initial density only makes sense if one has solved for $\delta_\chi(\vk)$ iteratively; otherwise at linear order one has done nothing. 

\subsection{Perturbative results}

In this section we will write down the expression for the reconstructed field in perturbation theory. We will first assume that the nonlinear density can be expressed as:
\beq
\delta(\vk) = \sum_{n=1}^{\infty} F(\vp_1,\cdots ,\vp_n)  \delta_0(\vp_1)\cdots  \delta_0(\vp_n). 
\eeq
We will need the first few kernels
\bea
F(\vp_1) &=& 1\nonumber \\
F(\vp_1,\vp_2) &=& l_2(\vp_1,\vp_2) +  F_Z(\vp_1,\vp_2)  \nonumber \\
F(\vp_1,\vp_2,\vp_3) &=&  l_3(\vp_1,\vp_2,\vp_3) +   2  [l_2(\vp_1,\vp_2) F_Z(\vp_1+\vp_2,\vp_3)]_{sym} + F_Z(\vp_1,\vp_2,\vp_3)\nonumber \\
 \kappa_2(\vp_1,\vp_2)&=& 1-{(\vp_1\cdot\vp_2)^2 \over \vp_1\cdot\vp_1 \vp_2\cdot\vp_2 } \label{eq:kappa2} \\
 l_2(\vp_1,\vp_2)&=& {3\over 14} \kappa_2(\vp_1,\vp_2) \nonumber \\
 \kappa_3(\vp_1,\vp_2,\vp_3) &=& {(\vp_1\cdot(\vp_2\times\vp_3))^2 \over \vp_1\cdot\vp_1 \vp_2\cdot\vp_2 \vp_3\cdot\vp_3 }  \nonumber \\
 l_3(\vp_1,\vp_2,\vp_3)&=& {1\over 6}\left(- {1\over 3} \kappa_3(\vp_1,\vp_2,\vp_3) + {5\over 21} [\kappa_2(\vp_1,\vp_2) \kappa_2(\vp_1+\vp_2,\vp_3)]_{sym}  \right)
\eea

We will use these expressions to obtain a formula for the reconstructed fields in terms of the initial condition $\delta_0$, 
\bea
\hat \delta_0^{S}(\vk) &=& \sum_{n=1}^{\infty} F_S(\vp_1,\cdots ,\vp_n)  \delta_0(\vp_1)\cdots  \delta_0(\vp_n) \nonumber \\
\hat \delta_0^{N}(\vk) &=& \sum_{n=1}^{\infty} F_N(\vp_1,\cdots ,\vp_n)  \delta_0(\vp_1)\cdots  \delta_0(\vp_n) 
\eea

In standard reconstruction
\beq
\hat \delta_0^{S}(\hat\vq)= \delta(\vx(\hat\vq))  J(\hat\vq)  \equiv \tilde \delta(\hat\vq) J(\hat\vq) \eeq
We can define:
\bea
\label{eq:ExpandDeltaChiInDelta0}
\delta_\chi(\vk) &=& \sum_{n=1}^{\infty} F_\chi(\vp_1,\cdots ,\vp_n)  \delta_0(\vp_1)\cdots  \delta_0(\vp_n) \nonumber \\
\tilde \delta(\vk) &=& \sum_{n=1}^{\infty} \tilde F(\vp_1,\cdots ,\vp_n)  \delta_0(\vp_1)\cdots  \delta_0(\vp_n) \nonumber \\
J(\vk) &=& \sum_{n=1}^{\infty} F_J(\vp_1,\cdots ,\vp_n)  \delta_0(\vp_1)\cdots  \delta_0(\vp_n),
\eea
in terms of which we find:
\bea
F_S(\vp_1) &=& \tilde F(\vp_1) \nonumber \\
F_S(\vp_1,\vp_2) &=& \tilde F(\vp_1,\vp_2) + [\tilde F(\vp_1)\tilde F_J(\vp_2)]_{sym} \nonumber \\
F_S(\vp_1,\vp_2,\vp_3) &=& \tilde F(\vp_1,\vp_2,\vp_3) + [\tilde F(\vp_1,\vp_2)\tilde F_J(\vp_3)+ \tilde F(\vp_1)\tilde F_J(\vp_2,\vp_3)]_{sym}. 
\eea
The solution for $\tilde\delta$ reads:
\bea
\tilde F(\vp_1) &=& 1 \nonumber \\
\tilde F(\vp_1,\vp_2) &=& F(\vp_1,\vp_2) -  [u(\vp_1,\vp_2) \tilde F(\vp_1) F_\chi(\vp_2)]_{sym} \nonumber \\
\tilde F(\vp_1,\vp_2,\vp_3) &=& F(\vp_1,\vp_2,\vp_3)  -  [u(\vp_1+\vp_2,\vp_3)  \tilde F(\vp_1,\vp_2) F_\chi(\vp_3)]_{sym} -  [u(\vp_3,\vp_1+\vp_2)  \tilde F(\vp_3) F_\chi(\vp_1,\vp_2)]_{sym}
\nonumber \\ &+& {1\over 2}[u(\vp_1,\vp_2)u(\vp_1,\vp_3)  F(\vp_1) F_\chi(\vp_2)F_\chi(\vp_3) ]_{sym} \nonumber \\
u(\vp_1,\vp_2)&\equiv & {\vp_1\cdot \vp_2 \over \vp_2\cdot \vp_2}.
\eea
For $J$ (which we only need to second order)  we have:
\bea
F_J(\vp_1) &=& -  F_\chi(\vp_1) \nonumber \\
F_J(\vp_1,\vp_2) &=&-F_\chi(\vp_1,\vp_2) + {1\over 2} \kappa(\vp_1,\vp_2) F_\chi(\vp_1)  F_\chi(\vp_2).  
\eea
Finally for $\delta_\chi$ we have:
\bea
\label{Fchi1}
F_\chi(\vp_1) &=& W(\vp_1) \nonumber \\
F_\chi(\vp_1,\vp_2) &=& W(\vp_1+\vp_2) F(\vp_1,\vp_2) -  [F_Z (\vp_1,\vp_2) F_\chi(\vp_1) F_\chi(\vp_2)] \nonumber \\
F_\chi(\vp_1,\vp_2,\vp_3) &=& W(\vp_1+\vp_2+\vp_3)  F(\vp_1,\vp_2,\vp_3)  -  [2 F_Z(\vp_1+\vp_2,\vp_3)  \tilde F_\chi(\vp_1,\vp_2) F_\chi(\vp_3)]_{sym} 
\nonumber \\ &-& F_Z(\vp_1,\vp_2,\vp_3)  F_\chi(\vp_1) F_\chi(\vp_2)F_\chi(\vp_3) .
\eea

The easiest way to get these formulas is to simply equate
\beq
\delta_Z[\delta_\chi] =   \sum_{n=1}^{\infty} F_Z(\vp_1,\cdots ,\vp_n)  \delta_\chi(\vp_1)\cdots  \delta_\chi(\vp_n) = W \delta = W  \sum_{n=1}^{\infty} F(\vp_1,\cdots ,\vp_n)  \delta_0(\vp_1)\cdots  \delta_0(\vp_n)
\eeq
and solve for $\delta_\chi$. 

One can easily use the above formulas to infer some properties of the reconstructed field. For example we can look at the quadratic kernels:
\bea
F_S(\vp_1,\vp_2) &=&  l_2(\vp_1,\vp_2) + F_Z (\vp_1,\vp_2) - \left[{(\vp_1+\vp_2)\cdot\vp_2 \over \vp_2\cdot\vp_2} W(\vp_2)\right]_{sym} \nonumber \\
F_\chi(\vp_1,\vp_2)/ W(\vp_1+\vp_2) &=&  l_2(\vp_1,\vp_2) + F_Z (\vp_1,\vp_2) \left[1-{W(\vp_1) W(\vp_2) \over  W(\vp_1+\vp_2)}\right].
\eea
It is easy to show that in the regime when $W\approx 1$ all the shift terms are canceled from both formulas; in fact,
\bea
F_S(\vp_1,\vp_2) &\approx& - {2\over 7}  \kappa_2(\vp_1,\vp_2) \nonumber \\
F_\chi(\vp_1,\vp_2)/ W(\vp_1+\vp_2) &\approx&  {3\over 14}  \kappa_2(\vp_1,\vp_2).
\eea
Indeed, this is true for the higher order kernels as well; the shift term are cancelled when $W\rightarrow 1$. 
Furthermore in the limit in which $p_1,p_2\gg k$ both kernels have a nice UV limit; they scale as $(k/p)^2$. In fact, in this limit
\bea
F_S(\vp_1,\vp_2) &\approx& {3 k^2\over 14 p^2}  \left(1 - {10\over 3} \mu^2 + {10\over 3} W(p) + {14\over 3} \mu^2 W(p)- {7\over 3} \mu^2 p W^\prime(p)\right) \nonumber \\
\label{eq:Fchi2limit}
F_\chi(\vp_1,\vp_2)/ W(\vp_1+\vp_2) &\approx&  {3 k^2\over 14 p^2}  \left(1 - {10\over 3} \mu^2 + {10\over 3} \mu^2 W(p)\right),
\eea
with $\mu=\vp\cdot\vk/p k$. Thus the kernels have the correct UV limit regardless of the window. This is again still true for the higher order kernels. 
In fact, both reconstruction formulas look very much the same, they only differ in terms proportional to the gravitational interaction. {From the fact that the second stage of the new and standard reconstructions differ only slightly in perturbation theory,  we expect that \emph{for a given displacement}, the second stage of the new and standard reconstruction should perform similarly. Therefore it makes sense that the extended standard reconstruction in \app{ExtendedStdRec}, which is using the eight-step displacement field $\vchi^{(8)}$, performs roughly as well as the new $\mathcal{O}(1)$ reconstruction based on the same displacement field (see \fig{1mr2Steps}). The key improvement from the new reconstruction comes from having a better displacement field (the second order correction and the reduced coefficients of the growth and tidal term play a much smaller role).}

One could improve the reconstruction by removing the quadratic and cubic pieces. At one loop the cubic piece could be incorporated in a transfer function on the linear field. Thus one could estimate the initial density using
\beq
\hat \delta_0 = t_1(k) \delta_\chi(\vk) + t_2(k) \kappa_2(\vp_1,\vp_2) t_1(\vp_1)\delta_\chi(\vp_1) t_1(\vp_2)\delta_\chi(\vp_2), 
\eeq
and choose $t_1$ and $t_2$ to minimize the difference with $\delta_0$ in simulations as described in \app{TransferFcns}. 
We tested this approach in the main text. 
 
\end{widetext}

\bibliography{draft_combined}

\label{lastpage}

\end{document}